\documentclass[preprint,superscriptaddress,nofootinbib,preprintnumbers,amsmath,amssymb]{revtex4-1}

\usepackage{amsfonts}
\usepackage{mathrsfs}
\usepackage{leftidx}
\usepackage{amssymb}
\usepackage{placeins}
\usepackage{relsize}
\usepackage{slashed}
\usepackage{multirow}
\usepackage[dvipsnames]{xcolor}

\usepackage[colorlinks]{hyperref}
\usepackage{epsfig}
\usepackage{graphicx}               
\usepackage{url}
\usepackage{float}

\newcommand{\be}{\begin{equation}}
\newcommand{\ee}{\end{equation}}
\newcommand{\bea}{\begin{eqnarray}}
\newcommand{\eea}{\end{eqnarray}}
\newcommand{\beq}{\begin{eqnarray}}
\newcommand{\eeq}{\end{eqnarray}}
\def\bit{\begin{itemize}}
\def\eit{\end{itemize}}
\def\ben{\begin{enumerate}}
\def\een{\end{enumerate}}

\newcommand{\Eq}[1]{Eq.~(\ref{#1})}

\newcommand\DN[1][\relax]{%
\ifx\relax#1\relax\else{}^{#1}\fi \!X}

\newcommand{\BE}{\textrm{BE}}
\newcommand{\g}{g_\text{dof}}

\begin{document}


\title{Nuclear Structure of Bound States of Asymmetric Dark Matter}
\author{Moira I. Gresham}
\affiliation{Whitman College, Walla Walla, WA 99362}
\author{Hou Keong Lou}
\affiliation{Theoretical Physics Group, Lawrence Berkeley National Laboratory, Berkeley, CA 94720}
\affiliation{Berkeley Center for Theoretical Physics, University of California, Berkeley, CA 94720}
\author{Kathryn M. Zurek}
\affiliation{Theoretical Physics Group, Lawrence Berkeley National Laboratory, Berkeley, CA 94720}
\affiliation{Berkeley Center for Theoretical Physics, University of California, Berkeley, CA 94720}

\begin{abstract} 

Models of Asymmetric Dark Matter (ADM) with a sufficiently attractive and long-range force gives rise to stable bound objects, analogous to nuclei in the Standard Model, called nuggets.  We study the properties of these nuggets and compute their profiles and binding energies.  Our approach, applicable to both elementary and composite fermionic ADM, utilizes relativistic mean field theory, and allows a more systematic computation of nugget properties, over a wider range of sizes and force mediator masses, compared to previous literature. We identify three separate regimes of nugget property behavior corresponding to (1) non-relativistic and (2) relativistic constituents in a Coulomb-like limit, and (3) saturation in an anti-Coulomb limit when the nuggets are large compared to the force range. We provide analytical descriptions for nuggets in each regime. Through numerical calculations, we are able to confirm our analytic descriptions and also obtain smooth transitions for the nugget profiles between all three regimes. We also find that over a wide range of parameter space, the binding energy in the saturation limit is an ${\cal O}(1)$ fraction of the constituent's mass, significantly larger than expectations in the non-relativistic case.  In a companion paper, we apply our results to synthesis of ADM nuggets in the early Universe.
\end{abstract}

\maketitle
\tableofcontents

\section{Introduction}

The last decade has seen a radical shift in the dominant paradigm for dark matter.  The weakly interacting massive particle (WIMP) paradigm, where the dark matter is a single stable and weakly interacting particle, is being surpassed by a wider view, where the dark matter is part of a larger dark sector.   These dark sectors generically feature dark forces, strongly or weakly coupled, and may contain dynamics that shape its behavior throughout the history of the Universe. The dark matter itself may either be a fundamental state, or a composite particle. As a consequence of the new dark force dynamics, the dark matter mass range is greatly enlarged, with masses from a keV to well above the weak scale.    

The implications for the dark sector---cosmologically, astrophysically and experimentally---are far-reaching.  The dark matter is generically self-interacting, implying changes in the structure of halos from dwarf galaxies to clusters of galaxies (see {\em e.g.} \cite{Spergel:1999mh,BoylanKolchin:2011de,Tulin:2013teo,Tulin:2017ara}).  Unlike the WIMP paradigm, the dynamics of the dark sector often do not freeze out early in the Universe.  Instead dark sector interactions continue to shape the evolution of our Universe.  In addition, the mechanisms for setting the relic abundance are many fold, from utilizing a particle asymmetry as in Asymmetric Dark Matter (ADM), to freeze-in and freeze-out and decay.

In this paper, we explore the structure of large bound states of DM, which we refer to as ``nuggets'' \cite{Wise:2014ola}.  Nuggets, like nuclei in the Standard Model, arise in the presence of an attractive dark force, and where the dark matter density is asymmetric such that particle and anti-particle do not annihilate when bound together.  As long as the mediator is sufficiently light, such that cold fusion is possible, large nuggets can be synthesized efficiently in the early Universe \cite{Wise:2014jva,Hardy:2014mqa,synthesis}.

These nuggets are a qualitatively different kind of DM candidate, with an impact on DM cosmology, constraints and search techniques, from the early Universe until today.  First, as we study in a companion paper, when nuggets are synthesized in the early Universe, their size distribution is typically very broad  \cite{synthesis}; many DM particles remain in small-sized bound states, while some nuggets can be very large and easily have masses close to the GUT scale or beyond. Second, the presence of an attractive force implies an impact on structure formation from DM self interactions.  Third, dark disks and dark stars may form, leading to new observational signatures late in the Universe.  Lastly, detection of these objects directly in a laboratory setting implies a diverse range of signatures from small bound states to very large ones. 

A UV complete model to describe nugget formation and evolution opens the ability to gain a unified and consistent understanding of the cosmology, constraints and relevant search techniques for such DM from the early Universe until today.   We employ a simplified model featuring fermionic DM with both a scalar and vector mediator of DM self interactions.  While our simplified model features fundamental degrees of freedom, we emphasize that it can describe the phenomenology of both elementary and composite DM.  In particular, such an elementary model, under the name of the ``$\sigma$-$\omega$'' \footnote{$\sigma$ ($\omega$) refers to the scalar (vector) meson mediating spin- and isospin-independent attractive (repulsive) forces among nucleons in SM nuclei. These mesons are also known as $f_0(500)$ and $\omega(782)$.}, or the Walecka model \cite{Walecka:1974qa}, has been shown to describe many of the bulk properties of SM nuclei, including density and binding energy.  Our goal, in a series of companion papers, is to study each aspect of the properties and cosmology of large bound states of ADM, from early Universe synthesis through structure formation.

The first step of this journey is to map the UV complete model onto the IR properties of the DM nuggets.  
We use relativistic mean field theory (RMFT) and existing techniques from nuclear physics to solve the Walecka model and obtain the structure of the nuggets over a range of mediator masses, accounting for the presence of both attractive and repulsive forces.   We note that Refs.~\cite{Wise:2014ola,Wise:2014jva} also studied the properties (binding energy and density profile) of nuggets in the weakly coupled Coulombic limit, where the mass of the force mediator can be neglected. They also assumed an ansatz for the fermi momentum profile in order to obtain a solution numerically, which potentially causes inaccuracy and prevents generalizations to other regimes of interest.  Utilizing techniques for solving the Walecka model in the context of nuclear physics, we are able to obtain solutions to the equations generally, for a wide range of force masses, and including both attractive and repulsive forces.  We reproduce the results of Refs.~\cite{Wise:2014ola,Wise:2014jva} in the relevant limit.

One important general feature of our results is that fermionic ADM bound through a scalar mediator can form very large, stable, nuggets that saturate at some (possibly quite large) size determined by the coupling and ratio of mediator and constituent masses. At saturation, the nugget number density and binding energy per constituent are constant as a function of size, similar to the behavior of SM nuclei. In this limit, the nuggets' constituents are relativistic and the attractive force is balanced by fermi pressure (and also repulsive forces if they are present). Saturation was not explicitly seen in \cite{Wise:2014ola}, as their weakly coupled descriptions become invalid for large bound states. We explicitly demonstrate the approach to saturation analytically and numerically, where the mediator mass becomes important and the attractive force becomes short ranged.  A second important feature is that, in substantial parts of parameter space, the binding energy is an ${\cal O}(1)$ fraction of the mass energy of the nugget, providing the possibility for large energy release during fusion. Beyond saturation, we derive approximate analytic formulae for non-saturation behaviors in the massless mediator limit, which also apply to the general massive mediator case before saturation. Solving the Walecka model numerically, we are able to confirm our analytic results, and also recover the Coulomb-like limit for weakly bound small nuggets.

The outline of this paper is as follows.  In Sec.~\ref{sec:model} we describe an elementary and composite model of DM and dark forces that form the basis for our analysis.  Then in Sec.~\ref{sec:nucDM}, we solve the equations of motion to obtain various physical properties for the ADM nuggets.  
In Sec.~\ref{sec:conclusion}, we conclude with an eye toward future work exploring the synthesis and impact of ADM nuggets on stellar and structure formation. In App.~\ref{app:numerics}, we detail the techniques used for solving the equations of motion, while highlighting numerical challenges and solutions.

\section{Models of Interacting Asymmetric Dark Matter}
\label{sec:model}
In this section we consider a class of viable models that can accommodate multi-DM bound states.
In order to have a substantial number of large ADM nuggets, the DM should carry a particle-anti-particle asymmetry. There are two natural classes of models to consider: an elementary model where the DM is a fundamental particle carrying a global symmetry, and a composite model where the DM is a dark baryon. We restrict ourselves to fermionic DM for simplicity.  
We consider a DM, $X$, interacting through a scalar mediator $\phi$ and vector mediator $V_\mu$, with a Lagrangian given by
\begin{align}
  \mathcal{L} = \bar X \left[i\slashed\partial - g_V\slashed V - (m_X -g_\phi \phi)\right]X +
\frac{1}{2}\left[ 
(\partial \phi)^2 - m_\phi^2\phi^2  - V(\phi)
\right] -\frac{1}{4}V_{\mu\nu}V^{\mu\nu} +\frac{1}{2}m_V^2 V_\mu^2\,.
\label{eq:lagrangian}
\end{align}
The presence of $\phi$ is necessary, as an attractive force is required for bound state formation. While \Eq{eq:lagrangian} contains many parameters, many salient features of dark nuclear physics can be obtained by restricting ourselves to specific examples. 
To this end, we will focus mostly on two cases: the elementary case where $V_\mu$ is absent, and a composite case where $m_V/m_X$ is relatively large. We will solve the model taking $V(\phi) = 0$ and then comment on the effect of adding $V(\phi)$ back in.

We do not address the source of asymmetry for the DM, which could come from other higher dimensional interactions \cite{Kaplan:2009ag}. It is convenient for the mediators to be light, such that the process $X+\bar X \rightarrow \phi \phi / V V$ can efficiently remove the thermal symmetric component of the DM through annihilations, allowing for efficient fusion at later times. This is analogous to the way that $e^+ e^- \rightarrow \gamma \gamma$ effectively depletes positrons in the early universe to leave only electrons, and so that hydrogen formation can proceed later on. In the elementary model, the scalar mediator can naturally arise from a dark Higgs mechanism that sets the mass scales of our Lagrangian.
The $\phi$s can either be cosmologically stable, or decay to either SM states ($e^+ e^-,~\gamma\gamma,~\nu \bar \nu$), and/or to other light hidden sector particles such as dark radiation.

This elementary model has a strongly coupled dual, where $X$ is interpreted as a baryon and $\phi$ ($V_\mu$) the scalar (vector) meson. 
 Given that the $\sigma\textrm{-}\omega$ model is relatively successful at describing bulk properties of SM nuclei, and that it is difficult to systematically include all possible composite states, we will use the Lagrangian in \Eq{eq:lagrangian} as a effective parameterization that captures the main qualitative features of DM nuggets. In the language of the composite model, $\phi$ is the lightest, parity-even and flavor-neutral scalar meson, and it functions as the dark glue binding dark composites together. The vector meson $V_\mu$ effectively mediates a repulsive interaction, but will generally be heavier and/or less strongly coupled than $\phi$, thus still allowing a net attractive force. The couplings are expected to be large and of $\mathcal{O}(1)$.  One notable omission is the pseudoscalar meson, but because of the axial nature of its interaction (which mediates spin-dependent interactions), it does not play a leading role in the large bound state limit. It is important to keep in mind the parameters in our composite model do not easily map onto fundamental parameters, and the validation of such a description requires comparisons with data and/or lattice simulation. The detailed spectroscopy of a general composite hidden sector can get very complicated and is beyond the scope of this paper. 

 For either the composite or fundamental model, we expect some large nugget states to have large spin; if the shell model for nuclei is any guide, up to order $N^{1/3}$ for some odd-$N$ ground states, where $N$ is the dark number of the nugget. Though large spin could impact the dynamical properties of nuggets, the bulk properties roughly scale as the volume, which allows us to ignore spin effects as a leading order approximation. 

\section{Nugget Properties}
\label{sec:nucDM}

We now turn to computing the characteristic size, density, and other physical properties of ADM nuggets utilizing relativistic mean field theory. 
 The physics of bound states can be very rich in general, as evidenced by the SM. For clarity, we will first consider the simplest scenario with an elementary $X$ and scalar mediator only; and for simplicity, we will set $V(\phi)=0$ in \Eq{eq:lagrangian}, though we will later parameterize the effects of nonzero $V(\phi)$. The simplification allows us to explore interesting features of nugget bound states without the complications of a large  parameter space. In Sec.~\ref{sec:few_body}, we consider small $N$ bound states and briefly discuss unique properties that may exist for specific $N$. In Sec.~\ref{sec:many_body} we study larger $N$ nuggets and their average properties, while ignoring $N$-specific features. In Sec.~\ref{sec:saturation} we provide analytic formulae for very large nuggets that have hit saturation and then discuss the effect of a scalar potential on saturation properties in Sec.~\ref{sec:potential}. Finally, in Sec.~\ref{sec:vector} we include the vector mediator, focusing on the composite scenario where $m_V$ is heavy, and discuss important differences from the scalar mediator only case. Throughout our discussions, the ADM nugget states are assumed to be in the ground state.

\subsection{Few Body Bound States}
\label{sec:few_body}
For bound states involving a small number of constituents, the overall nuclear density is not very large, and one typically does not expect the constituents to be relativistic as long as the interactions remain perturbative. The non-perturbative case requires more complicated calculations and we refer the reader to Ref.~\cite{Detmold:2014kba} for an example. In the weakly-coupled case, the wave functions of the DN can be obtained via a non-relativistic Schr\"odinger's equation. The case of two-body bound states has already been treated extensively in Refs.~\cite{PhysRevA.1.1577,Wise:2014jva}, and we summarize their main results here.  In order for a bound state to form, the range of the force, $m_\phi^{-1}$, should exceed the size of the wave function (typically set by the Bohr radius $r_B^{-1} =\alpha_\phi m_X/2$, where $\alpha_\phi \equiv g_\phi^2 /(4\pi)$).  More precisely, it has been shown that as long as $m_\phi^{-1}> 0.84 \, r_B^{-1}$,  a 1$s$ two-body bound state exists \cite{PhysRevA.1.1577}. In the small mediator mass limit, $m_\phi \ll \alpha_\phi m_X$, the force is close to Coulombic, and the two particle binding energy is simply $\BE_2 = {\alpha_\phi^2 m_X}/4$. The ground state has zero spin, which maintains the antisymmetry of the total wave function.

For $N>2$, Schr\"odinger's equation becomes highly nontrivial, and analytic solutions for the bound states are not available. One useful simplification, at moderate $N$, is to assume that on average, each individual constituent is under the influence of an emergent potential. This is the shell model, which enjoys phenomenological success in standard nuclear physics (see \cite{bertulani2007nuclear} for a detailed introduction). In the shell model, each individual constituent is treated as non-interacting, and the constituents simply fill up the available states from low to high energy according to Pauli's exclusion principle. There are typically many (approximately) degenerate eigenstates for the potential, which results in large energy gaps between specific states. This leads to local maxima in binding energy per constituent as a function of size, $N$; the sizes corresponding to local maxima are so-called ``magic numbers''. The existence of local maxima in binding energy per constituent can lead to the instability or metastability of nearby states, and is especially important to the story of nucleosynthesis. 

At larger $N$, the occurrence of magic numbers becomes sparse, and their effects become subdominant.  This is the case we study next.

\subsection{Many Body Bound States}
\label{sec:many_body}
When the number of constituents becomes large, one can employ relativistic mean field theory instead of computing many-body wave functions. This is the basic idea behind the $\sigma$-$\omega$ or Walecka model used to describe bulk properties of SM nuclei and nuclear matter in {\em e.g.}~neutron stars.\footnote{See \cite{Glendenning:1997wn} for a pedagogical introduction to the $\sigma$-$\omega$ model in the context of describing the equation of state for neutron stars.} Here we will largely follow the formalism and derivations presented for the Walecka model in Refs.~\cite{Walecka:1974qa, Negele:1986bp} and apply them to ADM nuggets; we refer the reader to these references for more detail. 

In the limit as the number of constituents becomes large, the ground state is expected to be approximately rotationally invariant. The occupancy for the bosonic fields is expected to be large, and can thus be treated classically. In particular, the scalar field is replaced by its expectation value. For the fermions, we have the equations of motion
\begin{align}
\left[i\slashed\partial - (m_X-g_\phi \phi(x))\right] X(x) =0,
\end{align}
where $\phi(x) = \langle \phi \rangle$ is a spatially varying classical field. 
The scalar field equation of motion is,
\begin{align}
\nabla^2 \phi  = m_\phi^2 \phi - g_\phi \langle \bar X X \rangle. \label{eq: phi eom}
\end{align}

Assuming that the variation of $\phi$ is small over the compton wavelength of $X$, $\phi$ then acts as a locally varying effective mass $m_*(x) \equiv m_X - g_\phi \phi$. At each spatial location, $X$ can then be treated as a non-interacting degenerate fermi gas with a locally constant mass $m_*(x)$. This is the Thomas-Fermi approximation, which has many applications to electronic many-body systems. The profile for $X$ is characterized by a local fermi density $k_F(x)$ (assuming zero temperature), and the energy for a given nugget profile is given by 
\begin{align}
  E(\phi(x), k_F(x)) = \int dr \, 4\pi r^2 \left\{
\frac{1}{2}\left[
(\nabla \phi)^2 + m_\phi^2 \phi^2 \right]
 + \frac{\g}{2\pi} \int_0^{k_F} d k\, k^2 \sqrt{k^2 + (m_X-g_\phi \phi)^2}
\right\}\,,
\label{eq:nugget_energy}
\end{align}
where $\g$ is the number of degrees of freedom for the fermion field ($\g=2$ for a single spin-1/2 Dirac fermion). In the ground state, the physical profiles $\phi(x)$ and $k_F(x)$ are those that minimize the energy functional for fixed dark number; this is the equilibrium condition. Assuming an abrupt cutoff for the fermi momentum ($k_F(r)=0$ for $r\ge R$) introduces an additional parameter, $R$, or the radius of the nugget. Variation with respect to $R$ will be proportional to $k_F(R)$, which vanishes and will be neglected. To minimize energy while holding dark number fixed, one introduces a Lagrange multiplier $\mu$, such that $\delta E - \mu \delta N = 0$. Physically, $\mu$ is the chemical potential, or the minimum energy change when an extra DM particle is added to a nugget. Using $N = \int d^3 \vec r \langle \bar X \gamma^0 X \rangle= \int d^3 \vec r \, \left( \g \int^{k_F(r)} { d^3 \vec k \over (2 \pi)^3} \right)=\frac{2\g}{3\pi}\int dr\, r^2\, k^3_F$, and given that \Eq{eq: phi eom} is equivalent to $\delta E/\delta \phi =0$, the equilibrium condition then reduces to 
\begin{align}
  \mu = \frac{\delta E /\delta k_F(r)}{\delta N/\delta k_F(r)} = \sqrt{k_F^2(r) + \left[m_X-g_\phi \phi(r)\right]^2}.
\label{eq:chemical_potential}
\end{align}
Note that $\mu$ does not have any spatial dependence, and as a result \Eq{eq:chemical_potential} yields a simple relationship between $k_F(r)$ and $\phi(r)$. In general $\mu$ still has a complicated dependence on $N$ and other parameters of the theory. 

\begin{figure}
\begin{minipage}{.45\textwidth}
\includegraphics[trim={1cm 3.5cm 1cm 1cm},clip,width=1.0\linewidth]{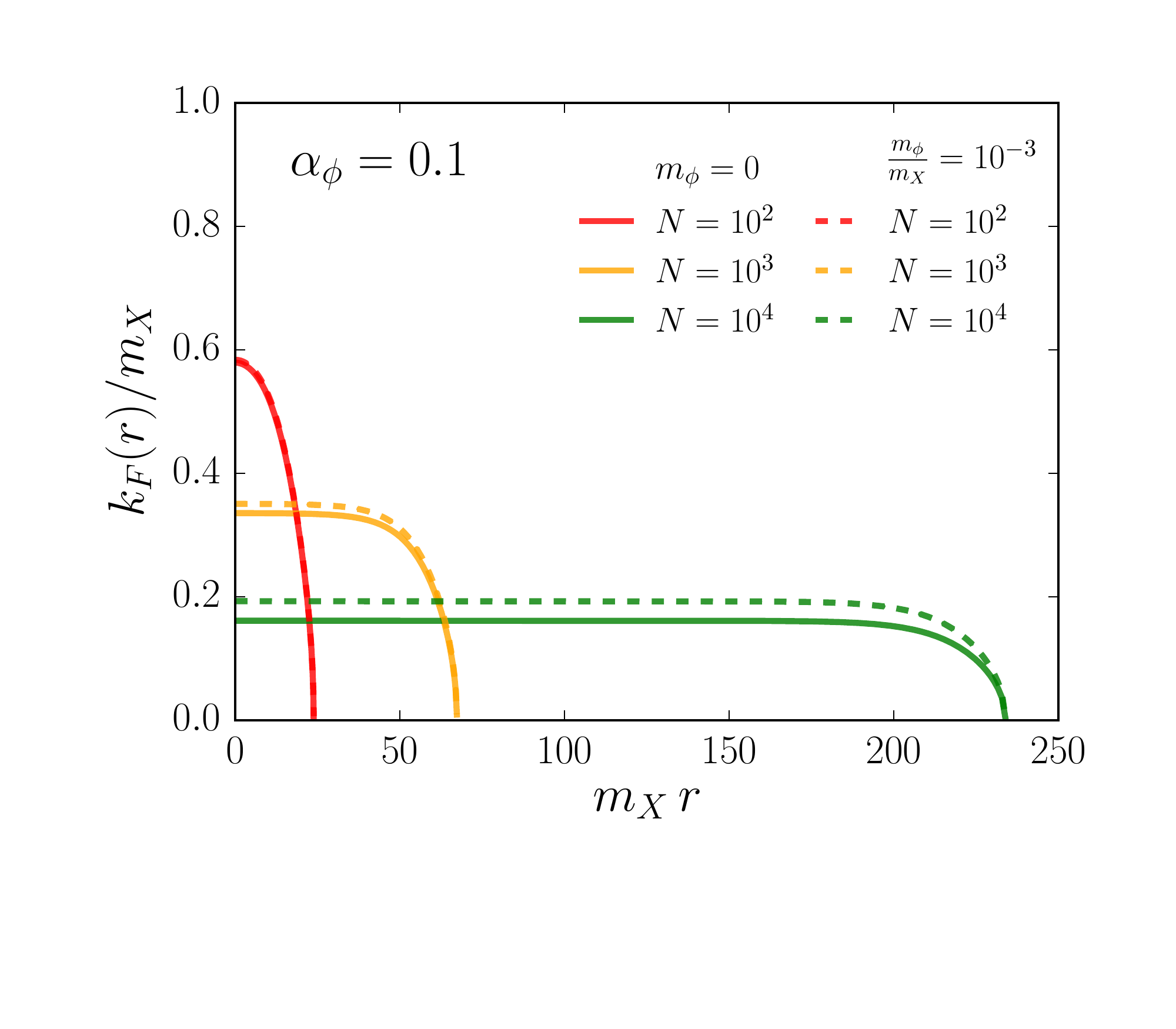}
\end{minipage}
\hspace{1cm}
\begin{minipage}{.45\textwidth}
\includegraphics[trim={1cm 3.5cm 1cm 1cm},clip,width=1.0\linewidth]{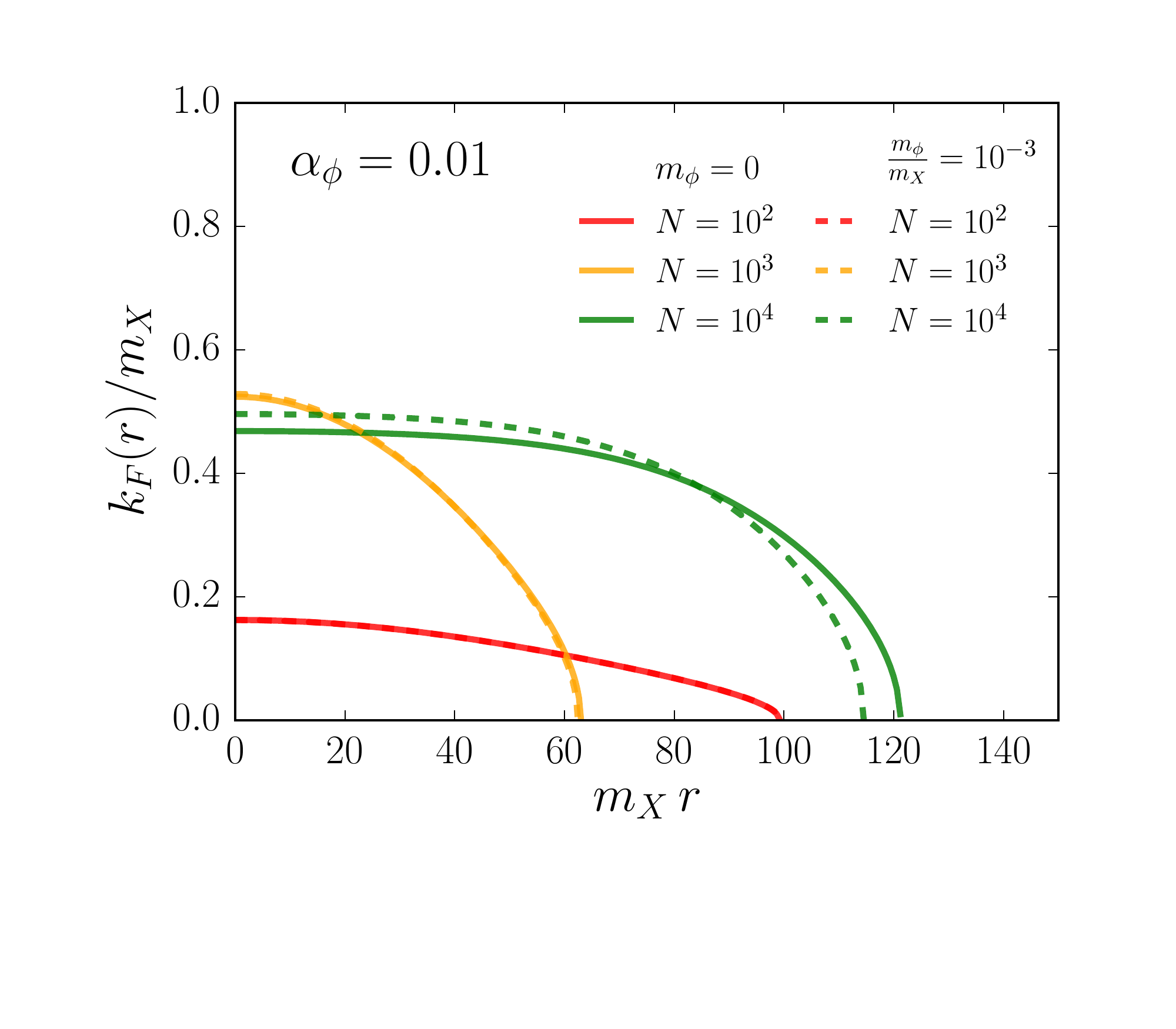}
\end{minipage}
\caption{Sample fermi-momentum profile, $k_F(r)$, for $\alpha_\phi=0.1$ (left) and $\alpha_\phi=0.01$ (right), and with varying $N\in (10^2, 10^3,10^4)$. The solid (dotted) line shows the profile for $m_\phi =0$ ($m_\phi = 10^{-3}m_X$). }
\label{fig:profile}
\end{figure}

Since the scalar density is given by $\langle \bar X X \rangle = \frac{\g}{2\pi}\int_0^{k_F(\phi)} dk\frac{k^2 (m_X-g_{\phi}\phi)}{\sqrt{k^2 + (m_X-g_{\phi}\phi)^2}}$ the scalar field equation of motion reads
\begin{align}
\nabla^2 \phi   = m_\phi^2 \phi -\frac{g_\phi\g}{2\pi}\int_0^{k_F(\phi)} dk\frac{k^2 (m_X-g_{\phi}\phi)}{\sqrt{k^2 + (m_X-g_{\phi}\phi)^2}},
\label{eq:nugget_ODE}
\end{align}
where $k_F$ can be written as a function of $\mu$ and $\phi$ using \Eq{eq:chemical_potential}. There are also additional boundary conditions for $\phi$, which is given by the requirement that it is continuous and differentiable at the boundary $r=R$, and that $\phi$ follows the equation of motion beyond the nugget boundary (i.e. $\phi(r)=\phi(R) e^{-m_\phi(r-R)} R/r$ for $r\ge R$). Together with the requirement that $\phi$ is well behaved at the origin, the boundary conditions are,
\begin{align}
  \partial_r\phi(0) = 0, \qquad g_\phi\phi(R) = m_X-\mu, \qquad 
  g_\phi\partial_r \phi(R) = (\mu - m_X)\frac{1+m_\phi R}{R}
\label{eq:nugget_bdy}.
\end{align}
In general, there are no closed form solutions and Eqs.~\ref{eq:nugget_ODE}-\ref{eq:nugget_bdy} must be solved numerically. Fig.~\ref{fig:profile} shows a few sample profiles for $\alpha_\phi \in \{0.1,0.01\}$ ($\alpha_\phi \equiv g_\phi^2/(4\pi)$) and $N\in \{10^{2},10^{3},10^{4}\}$, and for different scalar masses $m_\phi/m_X \in \{0, 10^{-3}\}$; we have also fixed $\g=2$.  The scalar mass, $m_\phi$, typically does not significantly impact nugget properties until $N$ becomes large, where, as $m_\phi$ increases, the nuggets become denser and smaller.

For a fixed $\alpha_\phi$ and $m_\phi$, there can be as many as three distinct regimes for the nugget profiles depending on $N$:
\begin{enumerate}
\item Small $N$: The nugget density is small and the constituents are largely non-relativistic. The density for the mediator remains relatively small, and the effective mass remains close to $m_X$. The nugget is small enough that $R\lesssim m_\phi^{-1}$, and the effect of the mediator mass, $m_\phi$, is insignificant. Using the non-relativistic formula for a fermi gas, and assuming a Coulomb-like potential, a behavior $R\simeq \sqrt[3]{81\pi^2/(4 N\g^2\alpha^3_\phi m^3_X)}$ can be derived (see \cite{Wise:2014jva} for details).
\item Medium $N$: The nugget is small enough, $R\lesssim m_\phi^{-1}$, that $m_\phi$ is largely unimportant, but the mediator density is large enough that $m_*$ is significantly different from $m_X$. The constituents become relativistic, leading to large fermi presssure that extends the nugget sizes. A scaling $R\sim N^{2/3}$ can be obtained (see text).
\item Large $N$ (Saturation): $R$ becomes much larger than $m_\phi^{-1}$. The binding energy, mediator density and $m_*$ all approach a constant. The nugget reaches a geometric limit where $R\sim N^{1/3}$.
\end{enumerate}

\begin{figure}
\begin{minipage}{.45\textwidth}
\includegraphics[trim={0cm 3.5cm 1cm 1cm},clip,width=1.0\linewidth]{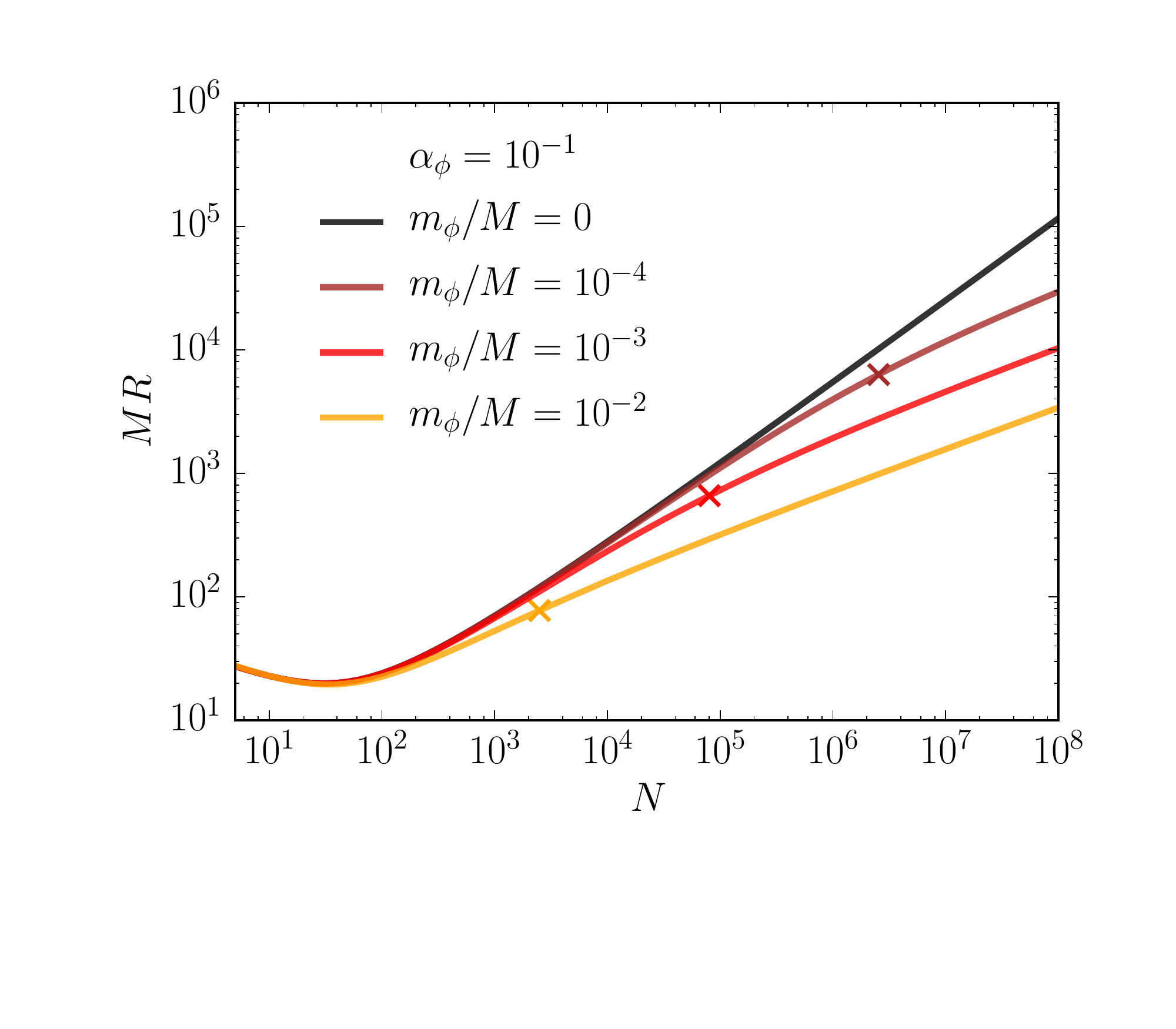}
\end{minipage}
\hspace{.5cm}
\begin{minipage}{.45\textwidth}
\includegraphics[trim={0cm 3.5cm 1cm 1cm},clip,width=1.0\linewidth]{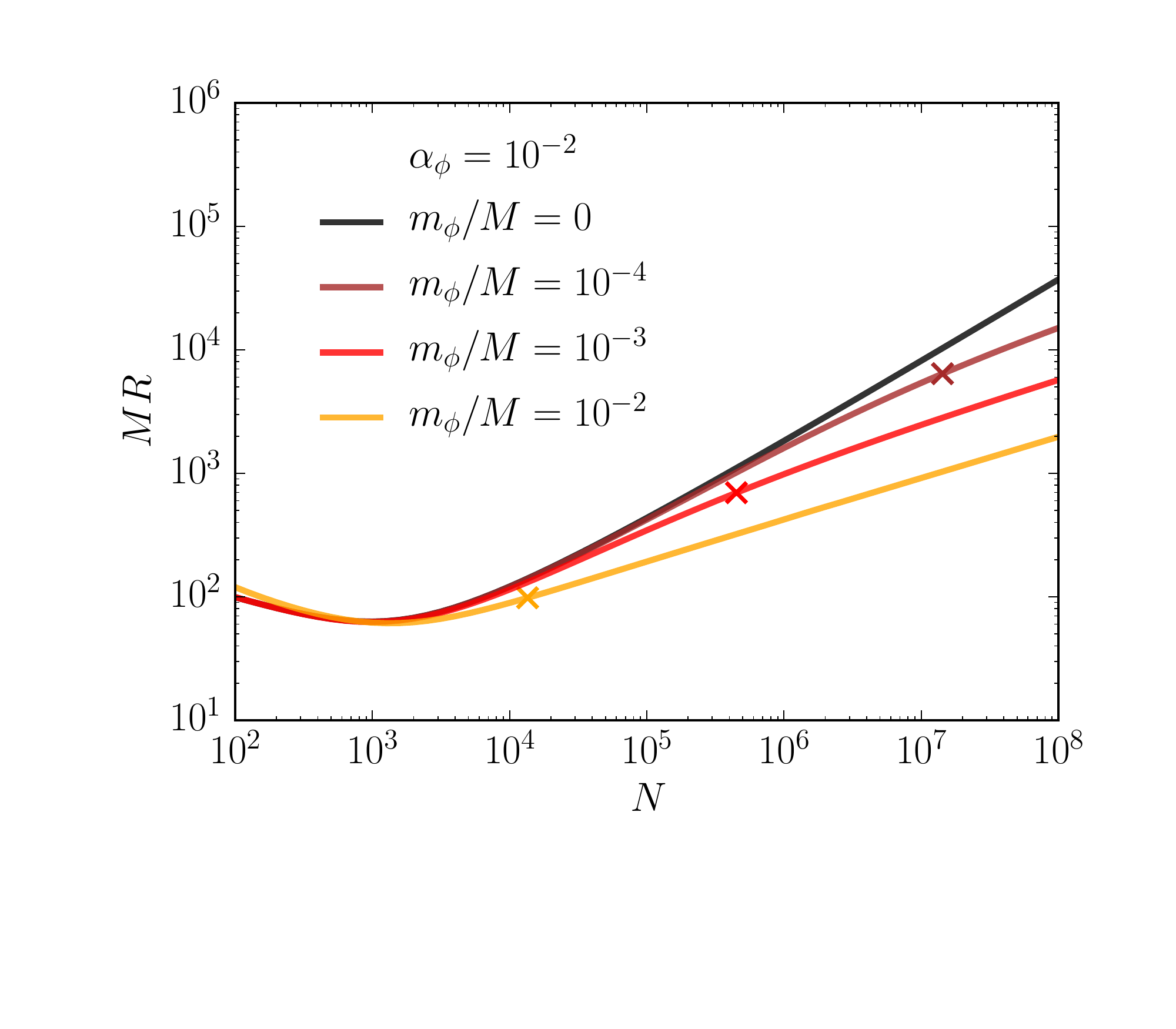}
\end{minipage}
\caption{Nugget radius versus $N$ for $\alpha_\phi=0.1$ (left) and $\alpha_\phi=0.01$ (right). At small $N$, $R \sim N^{-1/3}$, at moderate $N$, $R \sim N^{2/3}$, until saturation is reached (marked by an $\times$), where $R \sim N^{1/3}$.  See text for detailed discussion and derivation of the scalings.}
\label{fig:R_v_B}
\end{figure}

\begin{figure}
\begin{minipage}{.45\textwidth}
\includegraphics[trim={0.cm 3.5cm 1cm 1cm},clip,width=1.0\linewidth]{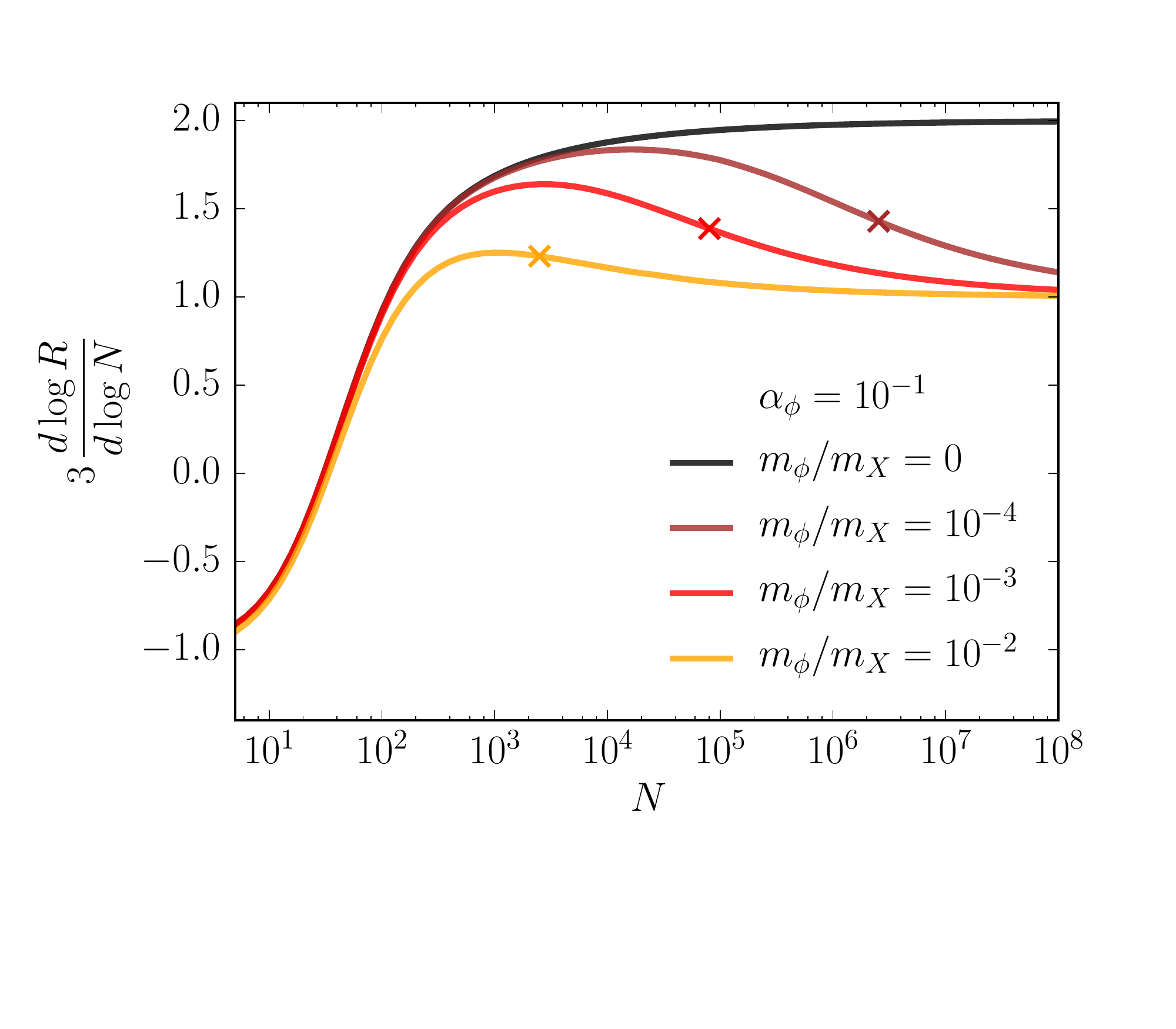}
\end{minipage}
\hspace{.5cm}
\begin{minipage}{.45\textwidth}
\includegraphics[trim={0.cm 3.5cm 1cm 1cm},clip,width=1.0\linewidth]{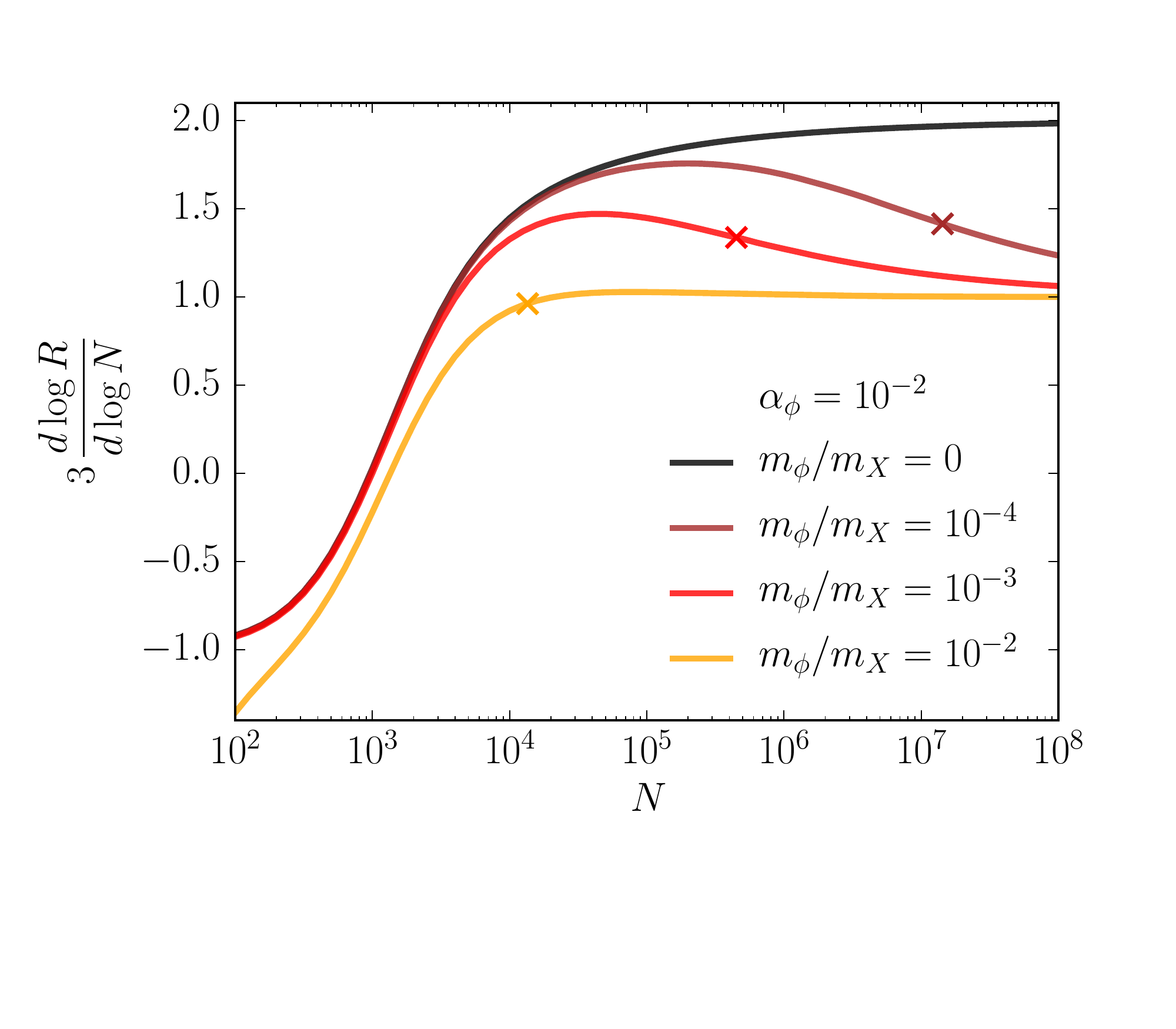}
\end{minipage}
\caption{Logarithmic derivative of the Nugget radius versus $N$ for different $\alpha_\phi=0.1$ (left) and $\alpha_\phi=0.01$ (right). The transition to saturation, defined as $N\ge \frac{4\pi}{3}\frac{n_\text{sat}}{m_\phi^3} $, is marked by an ``$\times$''. }
\label{fig:log_derivative_v_B}
\end{figure}

For a fixed $\alpha_\phi$ and $m_\phi$, one of the above regimes may be absent. The possibilities can be seen in Figs.~\ref{fig:R_v_B}-\ref{fig:log_derivative_v_B}, where $R(N)$ and its logarithmic derivative are shown for couplings $\alpha_\phi \in \{0.1, 0.01\}$ and mediator masses $m_\phi/m_X \in \{0, 10^{-4}, 10^{-3}, 10^{-2}\}$. For most of the benchmark cases, the scaling $N\sim R^{-1/3}$ at small $N$ is visible, though deviations from this scaling occur when the Coulomb-like approximation breaks down (when $r_B \sim (\alpha_\phi m_X)^{-1} \gtrsim m_\phi^{-1}$); this can be seen for the case $\alpha_\phi=0.01$ and $m_\phi/m_X =10^{-2}$.  In the massless limit, saturation is never reached, and the asymptotic behavior scales as $R\sim N^{2/3}$ . For light mediator masses $m_\phi/m_X \ll 1$, $R(N)$ follows the massless limit closely until $R \gtrsim m_\phi^{-1}$, when the nugget approaches saturation and the transition to a $R\sim N^{1/3}$ scaling occurs. 

\begin{figure}
\begin{minipage}{.45\textwidth}
\includegraphics[trim={0cm 3.5cm 1cm 1cm},clip,width=1.0\linewidth]{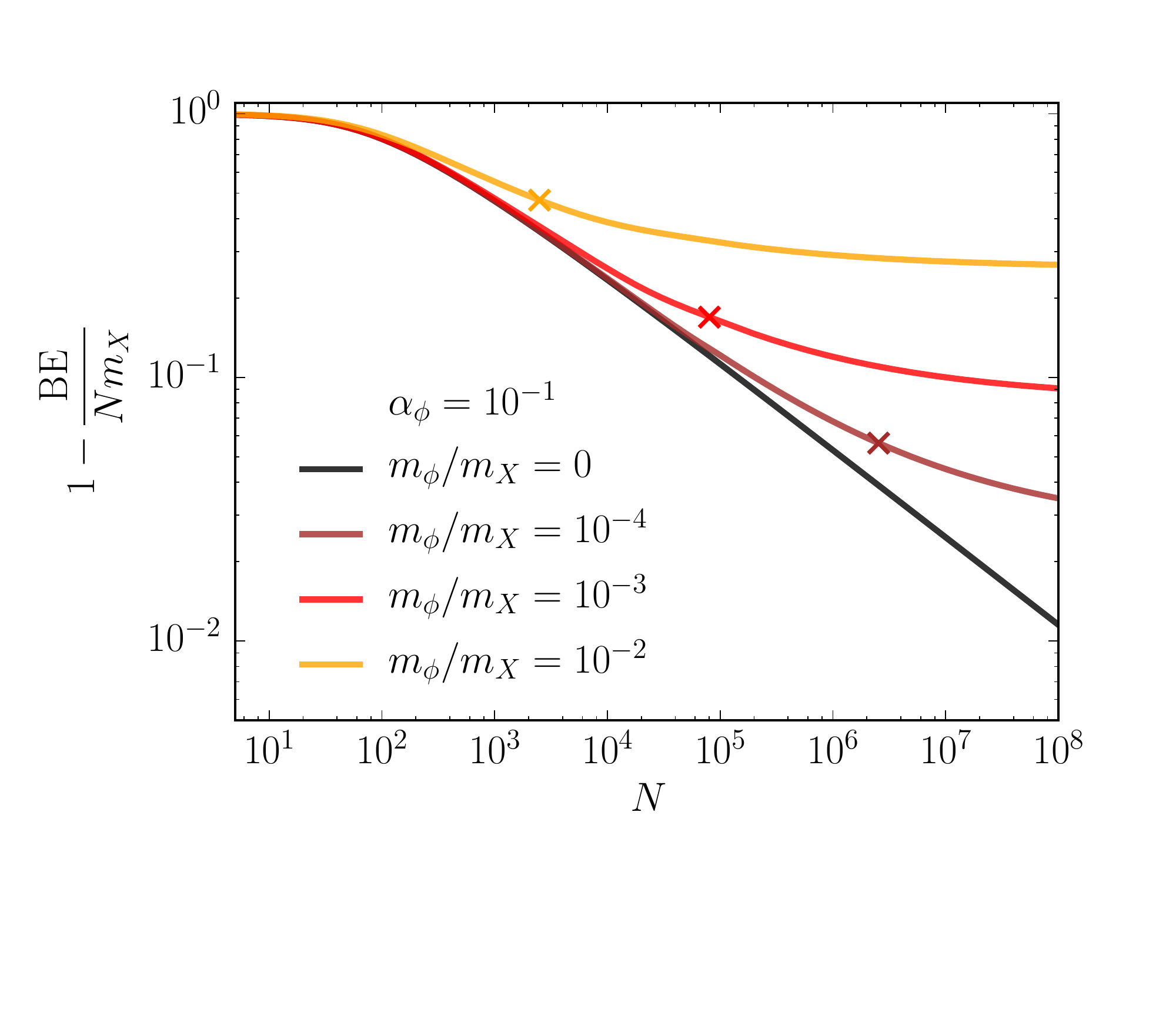}
\end{minipage}
\hspace{.5cm}
\begin{minipage}{.45\textwidth}
\includegraphics[trim={0cm 3.5cm 1cm 1cm},clip,width=1.0\linewidth]{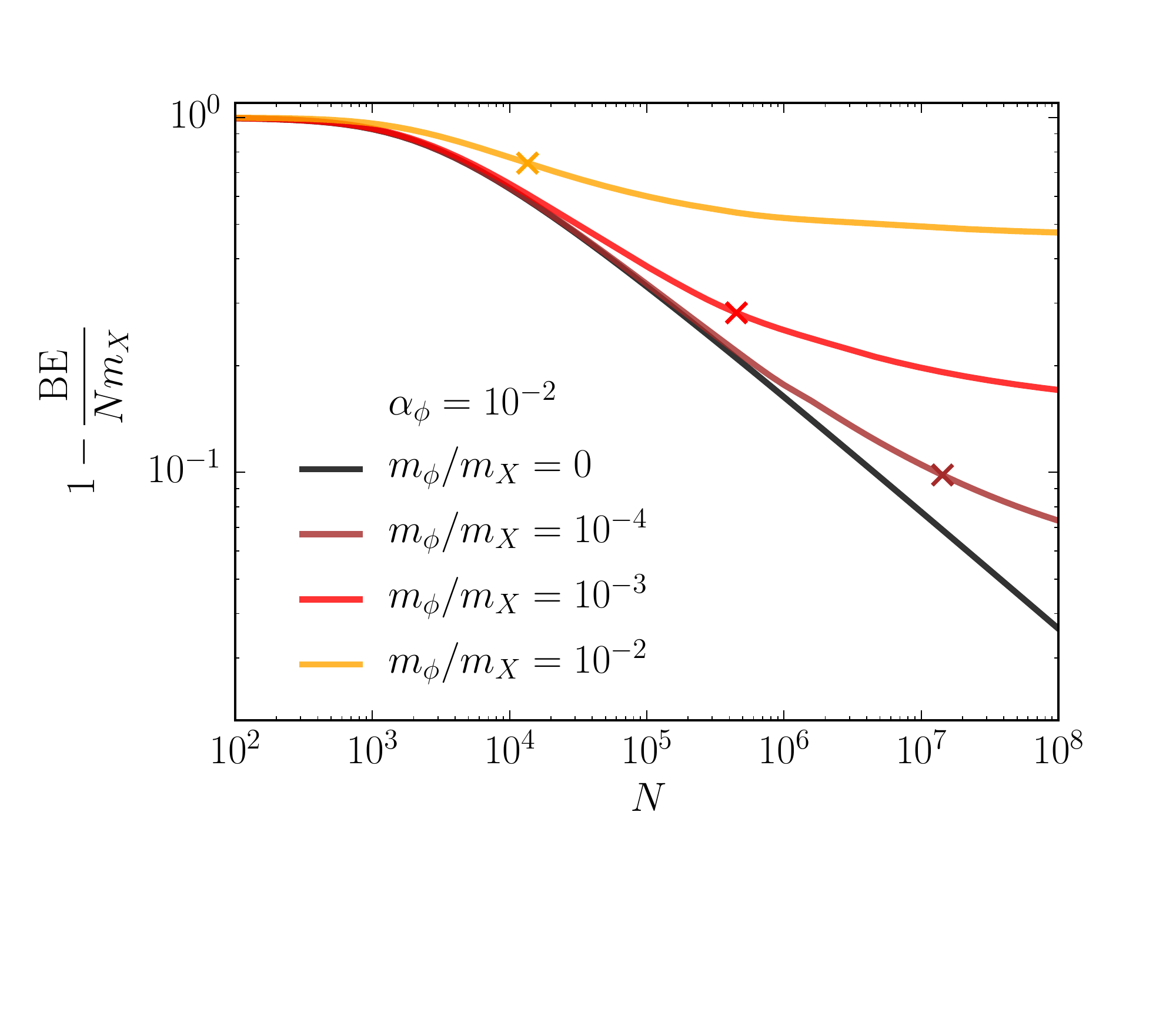}
\end{minipage}
\caption{Binding energy as a function of $N$ for $\alpha_\phi=0.1$ (left) and $\alpha_\phi=0.01$ (right). The transition to saturation, defined as $N\ge \frac{4\pi}{3}\frac{n_\text{sat}}{m_\phi^3} $, is marked by an ``$\times$''. }
\label{fig:BEvB}
\end{figure}

Fig.~\ref{fig:BEvB} shows the nugget binding energy per constituent as a function of the nugget number for the same benchmark points. For the massless case, the binding energy approaches the rest mass of the individual constituent, while for non-zero $m_\phi$, the binding energy approaches a constant at large $N$.  These figures also mark an estimate when saturation happens, which is defined by 
$N\gtrsim (r_0 m_\phi)^{-3}$ where $r_0$ is the saturation length scale; in particular $R = r_0 N^{1/3}$ and the number density is $n_\text{sat} =\left( {4 \over 3} \pi r_0^3 \right)^{-1}$ in the saturation limit. 
At large $N$, the binding energy is well described by a formula analogous to the semi-empirical mass formula for SM nuclei,
\begin{align}
M_N = N m_X -  \BE_N \simeq \mu_0 N + \epsilon_{\rm surf}{N^{2/3}} ,
\label{eq:energy_surface}
\end{align}
where $\mu_0$ is the bulk rest energy coefficient, equivalent to chemical potential energy in the infinite matter limit (as we will see), and $\epsilon_{\rm surf}$ is the surface term rest energy coefficient. Near saturation the nugget surface area scales as $N^{2/3}$; the surface term accounts for the lack of close-range interactions between constituents near the surface that would otherwise reduce the energy of the configuration. By fitting the curves in Fig.~\ref{fig:BEvB} to \Eq{eq:energy_surface} for $N > N_\text{sat}$, where
\beq
 N_\text{sat} \equiv (r_0 m_\phi)^{-3} \equiv {{4 \over 3} \pi n_\text{sat} / m_\phi^3} \label{eq:Nsat}
\eeq with the saturation length scale $r_0$ calculated in the $N \rightarrow \infty$ limit (see below),  we obtained the bulk and surface contribution to the binding energy for our benchmark cases, as shown in Table~\ref{tab:I}. In the next section, we obtain an analytic expression for the bulk contribution by examining the $N \rightarrow \infty$ (infinite matter) limit; the bulk parameters obtained through the fit match the analytic values to within less than $1\%$, indicating that our characterization of the ``saturation limit'', encapsulated in  \Eq{eq:energy_surface} and \Eq{eq:Nsat}, is self consistent.

\begin{table}[h]
  \centering
  \renewcommand{\arraystretch}{1.2}
  \setlength{\tabcolsep}{10pt}
  \begin{tabular}{c|c|c|c}
    $\alpha_\phi$ & $m_\phi/m_X$ & $\mu_0/m_X$ & $\!\!\!\epsilon_{\rm surf}/m_X\!\!\!\!\!$ \\
    \hline
    \multirow{3}{*}{$0.1$} & $10^{-2}$ & 0.26 & 3.2\\
    & $10^{-3}$ & 0.083 & 3.8\\
    & $10^{-4}$ & 0.026 & 4.2\\
    \hline
    \multirow{3}{*}{$0.01$} & $10^{-2}$ & 0.46 & 7.8\\
    & $10^{-3}$ & 0.15 & 11\\
    & $10^{-4}$ & 0.046 & 13\\
    \hline
  \end{tabular}
  \caption{\label{tab:I} Numerical values of $\mu_0$ and $\epsilon_{\rm surf}$ for our benchmark nugget models; these parameters are obtained by fitting \Eq{eq:energy_surface} to the curves in Fig.~\ref{fig:BEvB} at large $N>N_\text{sat}$.}
\end{table}

\subsection{Saturation}
\label{sec:saturation}

Thus far our discussion of the large $N$ behavior has been descriptive. However it is possible to obtain explicit analytic formulae in the large $R, N$ limit, as is standardly done within the $\sigma$-$\omega$ (Walecka) model in nuclear physics (see e.g.~\cite{Negele:1986bp} or \cite{1996cost.book.....G}). One can solve \Eq{eq:nugget_ODE} by simply replacing $k_F(x)$ and $\phi(x)$ with constants $k_{F0}$ and $\phi_0$, though an extra equation is needed to solve for the two unknowns. This can be obtained by considering the chemical potential, $\mu$, which should approach a constant as well, $\mu =dE/dN \rightarrow \mu_0$.  This in turn implies that $E = N \mu_0$, which serves as the second equation. Since pressure is $p = - \left( \partial E \over \partial V\right)\big|_N = (\mu - E/N)/V$, where $V$ is volume, the second (equilibrium) condition is equivalent to setting the pressure to zero.  These two equations can be recast in terms of the variables $m_* /m_X = 1 - g_\phi \phi_0/m_X$ and $k_F/m_X$, with the physical parameters entering only in the combination 
\beq C_\phi^2 = {2 \g \over 3 \pi} \alpha_\phi {m_X^2  \over  m_\phi^2}\,,
\eeq
such that
\begin{align}
1 - {m_* \over m_X} &= 3 C_\phi^2  \left( {m_* \over m_X} \right)\int_0^{\frac{k_{F0}}{m_X}} dx\frac{x^2}{\sqrt{x^2 + (m_*/m_X)^2}} \notag\\
p \left( {\g m_X^4 \over 6 \pi^2}\right)^{-1} &= - \frac{1}{
2 C_\phi^{2}}
\left(1 - {m_* \over m_X}\right)^2
 + \int_0^{\frac{k_{F0}}{m_X}} d x\, \frac{x^4}{ \sqrt{x^2 + (m_*/m_X)^2}}=0.
\label{eq:nugget_infinite_condition}
\end{align}

A solution to \Eq{eq:nugget_infinite_condition} with $p=0$ and with positive binding energy exists only for large enough $C_\phi^2$, implying a threshold for stability of (infinitely) large bound states.  We find the threshold to be  $ C_\phi^2 \gtrsim 1.1$. Larger $C_\phi^2$ corresponds to a larger attractive force, which requires a larger $k_{F0}/m_*$ to balance the total pressure. Since $m_*$ is the effective $X$ mass, large $k_{F0}/m_*$ also corresponds to effectively relativistic constituents. In the limit of large $k_{F0}/m_*$ one can show that \Eq{eq:nugget_infinite_condition} implies,
\begin{align}
{m_* \over m_X} & \rightarrow {1 \over 3}\sqrt{2 \over C_\phi^2},\\
{k_{F0} \over m_X} &\rightarrow \sqrt{\sqrt{2 \over C_\phi^2}\left(1 - {1 \over 3}\sqrt{2 \over C_\phi^2}\right)} . 
\end{align} Here it is apparent that $m_*$ falls more rapidly as $C_\phi$ grows than does $k_F$, meaning that the constituents are becoming more relativistic at saturation as $C_\phi$ grows even though the fermi momentum is simultaneously decreasing.  
Other properties can then be calculated in this limit. For example, $\left( {4 \over 3} \pi r_0^3\right)^{-1} = n_\text{sat} = {\g \over 6 \pi^2} k_{F0}^3$ so that $r_0 m_X = \left( {2 \g \over 9 \pi }\right)^{-1/3} \left( k_{F0} \over m_X \right)^{-1}$. Also, 
\beq
{\mu_0 \over m_X} = \left(1-{\text{BE}_N \over m_X N }\right) = \sqrt{\left({k_{F0} \over m_X}\right)^2 +\left({m_* \over m_X}\right)^2}. \label{eq:mu0}
\eeq
 The above expressions are quite accurate even at moderate values of $C_\phi^2$.
For $C_\phi^2 \gg 1$, 
\beq
{k_{F0} \over m_X} ,  {\mu_{0} \over m_X} \rightarrow 
\left( \frac{2}{C_\phi^2} \right)^{1/4} . \label{eq: kF in small mPhi limit}
\eeq

The saturation limit allows a simple geometric interpretation: the addition of constituent particles is analogous to adding liquid to an incompressible fluid. The binding energy per particle is simply $\BE_N /N = m_X - E/N = m_X-\mu$, and the nugget number scales directly as the volume of the nugget, {\em i.e.} $N= 4 \pi n_\text{sat} R^3/3$. 

The picture breaks down, however, when the mediator becomes massless, where the infinite volume limit also forces $n_\text{sat}\rightarrow 0$. In this case, the physics of large $R$ nuggets depends on a nonlinear differential equation. It is instructive to rewrite \Eq{eq:nugget_ODE}, by defining $f(r)=(m_X - g_\phi\phi(r))/\mu$, so that
\begin{align}
  \frac{d}{dr}
\left[
  \frac{1}{2}f'^2 - 4\pi\alpha_\phi \mu^2 \int_{0}^f \rho(y)dy + 
  m_\phi^2 \left( \frac{m_X}{\mu} f -\frac{1}{2}f^2\right)\right]
  =\frac{2}{r}f'^2
,
\qquad 
\rho(y)=\frac{\g }{2\pi^2}\int_0^{\sqrt{1-y^2}} \frac{y k^2\, dk}{\sqrt{k^2+y^2}}\,,
\label{eq:analytic}
\end{align}
with the boundary conditions: $f'(0)=0$, $f(R)=1$ and 
${f'(R)=(1+m_\phi R)(m_X -\mu)/(\mu R)}$. When $N$ is large, $f'$ is generally small until $r$ becomes large, such that one can ignore the term $f'^2/r$ and fully integrate the differential equation. Taking the limit $m_\phi = 0$, $f(0) \sim 0$ and $\mu \ll m_X$, one has the simple relations
\begin{align}
  \frac{1}{2}\left(\frac{m_X}{\mu R} \right)^2\simeq 4\pi \alpha_\phi \mu^2 \int_0^1 \rho(y)dy = \frac{\g \alpha_\phi \mu^2}{6\pi}\,.
\end{align}
Together with the approximation that $N\simeq \frac{2\g}{9\pi}R^3 k^3_F(0)\simeq \frac{2\g}{9\pi}R^3 \mu^3$, one obtains 
\begin{align}
R\simeq   \frac{N^{\frac{2}{3}}\sqrt{\alpha_\phi}}{m_X}\left(\frac{243\pi}{16 \g}\right)^{\frac{1}{6}}.
\end{align}
In the limit of finite $m_\phi$ and $R \gg m_\phi^{-1}$, the $R$ dependence in \Eq{eq:analytic} drops out as $f'(R)\rightarrow m_\phi(m_X -\mu)/\mu$.  Here one finds that \Eq{eq:nugget_infinite_condition}, the saturation limit, is recovered. 

\subsection{Inclusion of Scalar Potential Terms}
\label{sec:potential}

In this section we discuss the effects of additional scalar interactions on the properties of nuggets in the saturation limit. Specifically we consider a nonvanishing potential $V(\phi)= \lambda_4 {\g g_\phi^4 \over 4! \pi^2} \phi^4$ with $\lambda_4>0$ to maintain stability in the UV. The potential modifies the equilibrium condition and $\phi$ field equations, \Eq{eq:nugget_infinite_condition}, as follows:
\begin{align}
1 - {m_* \over m_X}  &= - C_\phi^2 \lambda_4 \left( 1 - {m_* \over m_X}\right)^3 +3 C_\phi^2  \left( {m_* \over m_X} \right)\int_0^{\frac{k_{F0}}{m_X}} dx\frac{x^2}{\sqrt{x^2 + (m_*/m_X)^2}} \notag\\
p \left( {\g m_X^4 \over 6 \pi^2}\right)^{-1} &= - \frac{1}{
2 C_\phi^{2}}\left(1 - {m_* \over m_X}\right)^2 - \frac{\lambda_4}{4}\left(1 - {m_* \over m_X}\right)^4
 + \int_0^{\frac{k_{F0}}{m_X}} d x\, \frac{x^4}{ \sqrt{x^2 + (m_*/m_X)^2}}=0.
\label{eq:nugget_infinite_condition_w_potential}
\end{align}
\Eq{eq:mu0} for the energy per constituent still holds but the equilibrium values for $k_{F0}$ and $m_*$ will of course change with the addition of the $\lambda_4$ term, according to \Eq{eq:nugget_infinite_condition_w_potential}. When $m_*/m_X$ is small, corresponding to large field values, $\phi$, we see that the quartic term is important when $\lambda_4 \gtrsim 1/C_\phi^2$. It is instructive to consider the case when $C_\phi^2 \ggg 1$ and $\lambda_4 \lesssim 1/C_\phi^2$. In this case, nuggets will still saturate in the relativistic limit, but the equilibrium values for $m_*$ and $k_F$ will change according to,
\begin{align}
{m_* \over m_X} & \rightarrow {1 \over 3}\sqrt{2 \over C_\phi^2} \left(1 +  \lambda_4 C_\phi^2 \right)\left( 1 + {\lambda_4 C_\phi^2 \over 2}\right)^{-1/2},\\
{k_{F0} \over m_X} &\rightarrow  \left({2 \over C_\phi^2}\right)^{1/4} \left( 1 + {\lambda_4 C_\phi^2 \over 2}\right)^{1/4}. \label{eq:mphi_w_potential}
\end{align}
The quartic term increases both $k_{F0}$ and $m_*$. This leads to an increase in energy per constituent (decrease in binding energy) and an increase in density. At the same time, note that $k_{F0}/m_*$ decreases, meaning the nuggets are less relativistic at saturation. This indicates that the net effect of the quartic term is to provide an effectively repulsive force. Once $\lambda_4 C_\phi^2$ is large enough so that $m_\phi \ll 1$ no longer holds (when ${{2 \sqrt{\lambda_4}} \over 3} \gtrsim 0.1$), our approximation breaks down. At the same time,  since increasing $m_*$ is equivalent to decreasing $\phi$, the quartic term is self-moderating in the sense that it forces saturation at lower values of $\phi$.

\begin{figure}
\begin{minipage}{.45\textwidth}
\includegraphics[trim={0cm 3.5cm 1cm 1cm},clip,width=1.0\linewidth]{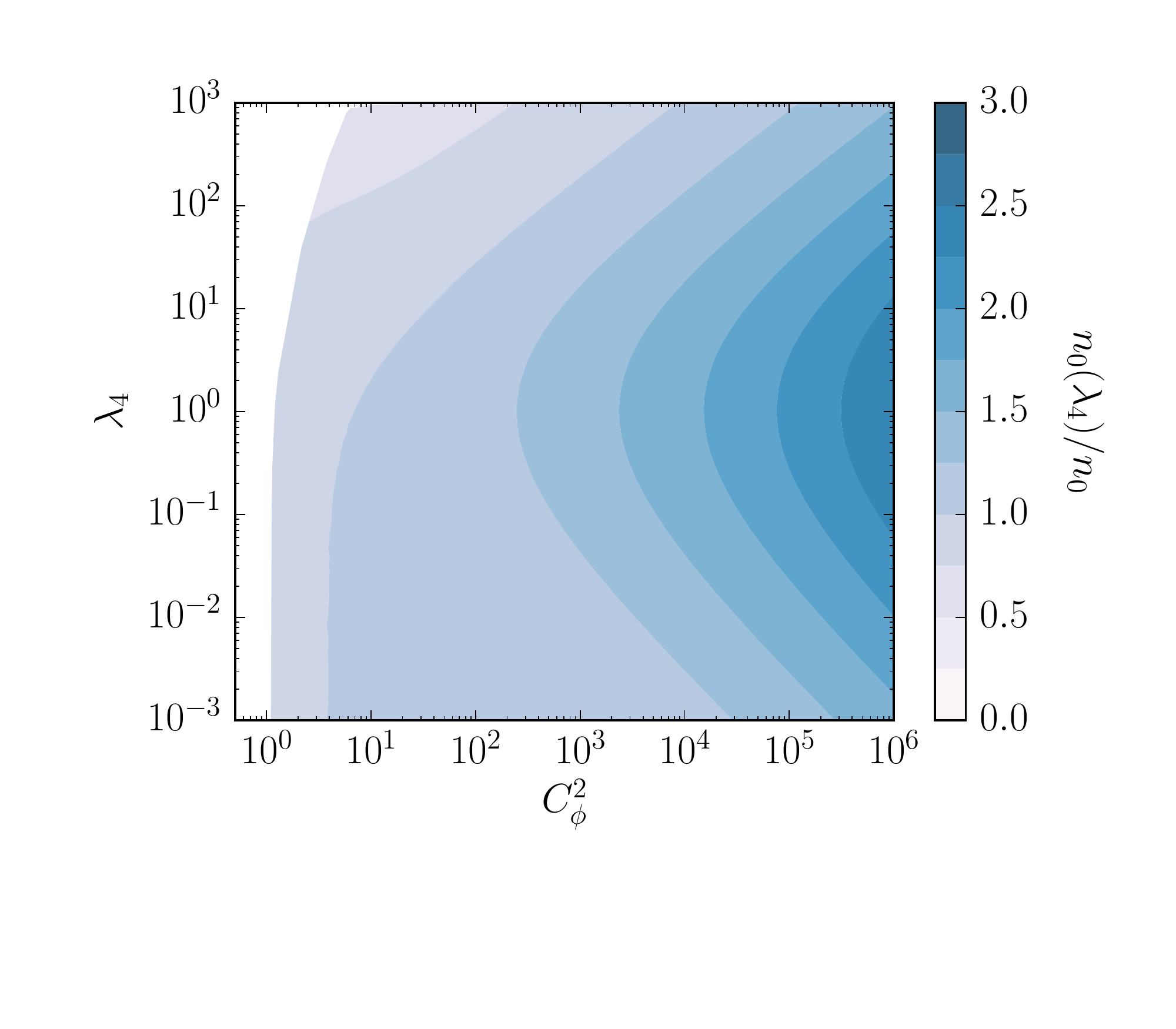}
\end{minipage}
\hspace{.5cm}
\begin{minipage}{.45\textwidth}
\includegraphics[trim={0cm 3.5cm 1cm 1cm},clip,width=1.0\linewidth]{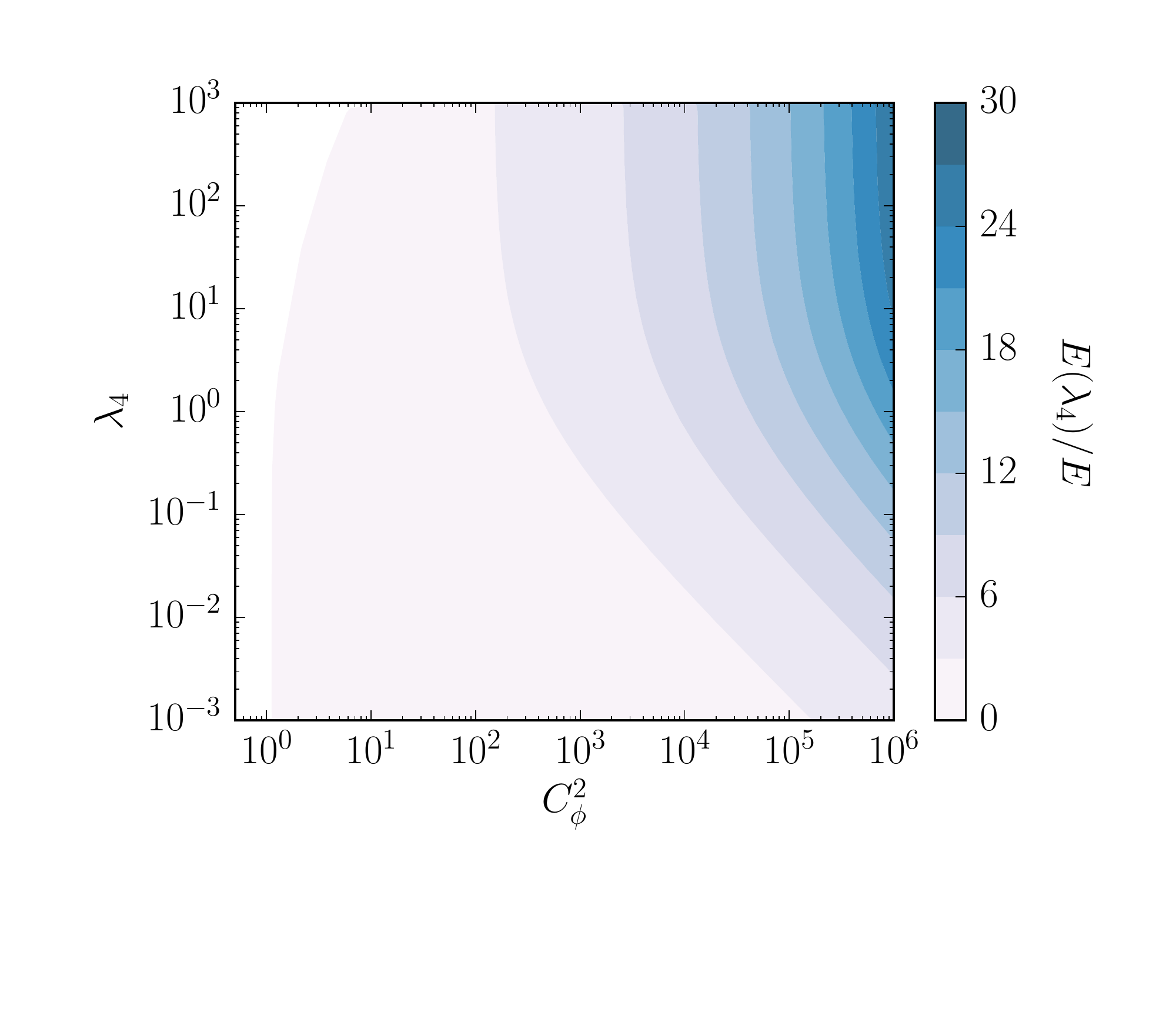}
\end{minipage}
\caption{The figure on the left (right) shows the nugget density (nugget energy density) for models with a non-vanishing quartic term $V(\phi)= \lambda_4 {\g g_\phi^4 \over 4! \pi^2} \phi^4$ as compared to a zero quartic term. The uncolored areas indicate regions where no saturation limit exists.}
\label{fig:phi4}
\end{figure}

Fig.~\ref{fig:phi4} shows the modification of the saturation number density (left) and energy density  (right) of the nuggets with the inclusion of the $\phi^4$ term. Note that the contours start curving around the point where ${{2 \sqrt{\lambda_4}} \over 3} \gtrsim 0.1$; for smaller values of $\lambda_4$, the estimate \Eq{eq:mphi_w_potential} should hold when $C_\phi^2 \gg 1$. In the white regions, no infinite bound matter limit exists.

The inclusion of a quadratic term changes the effective scalar mass, and therefore also the effective force range, according to 
\beq
m_{\phi \text{eff}} = m_\phi \sqrt{1 + 2 V(\langle \phi \rangle)/m_\phi^2 \langle \phi \rangle^2} = m_\phi \sqrt{1 + \lambda_4 C_\phi^2 (1 - m_*/m_X)^2 / 2}.
\eeq
With the inclusion of a potential, we expect a good estimate of the saturation size, $N_\text{sat}$, to be given as in \Eq{eq:Nsat} but with $m_\phi \rightarrow m_{\phi \text{eff}}$.

\subsection{Inclusion of Vector Mediator}
\label{sec:vector}

So far our treatment has been restricted to the scalar mediator, where the nuclear properties are controlled by a single function $k_F(r)$ or $g_\phi\phi(r)$. The introduction of a vector mediator leads to a repulsive force, and will generally lower the binding energy. In the static, classical limit, where the field is set to its expectation value and ignoring possible spin effects, only the temporal component of the classical vector field, $V_0$, is non-vanishing. The equation of motion for the classical vector field is then
\begin{align}
 - \nabla^2 V_0 + m_V^2 V_0
  = g_V \langle \bar X \gamma_0 X \rangle.
\label{eq:EOM_vector}
\end{align}
Analogous to electromagnetism, the right side of \Eq{eq:EOM_vector} is simply $g_V$ times the nugget density, $\langle \bar X \gamma_0 X \rangle \rightarrow \frac{\g k_F^3}{6 \pi^2} $. For positive coupling $g_V$, the resulting potential $V_0$ is always positive. The energy functional receives an additional contribution from the vector field,
\begin{align}
  E_V = \int dr\, 4\pi r^2 \;\left\{-\frac{1}{2}\left[(\nabla V_0)^2 + m_V^2 V_0^2 \right] + g_V V_0 \frac{\g k_F^3}{6\pi^2}\right\}\,,
\end{align} which can be seen to be manifestly positive through integrating by parts and employing the vector field equation of motion.
Carrying out the same computation as before with $\mu=dE/dN$ leads to a modified chemical potential
\begin{align}
\mu= g_V V_0 + 
   \sqrt{k_F^2 + (m_X - g_\phi\phi)^2}\,.
\end{align}
The presence of $V_0$ effectively increases the local chemical potential, and thus leads to a lower binding energy for the same nugget number. 

\begin{figure}
\begin{minipage}{.45\textwidth}
\includegraphics[trim={1cm 3.5cm 0cm 1cm},clip,width=1.0\linewidth]{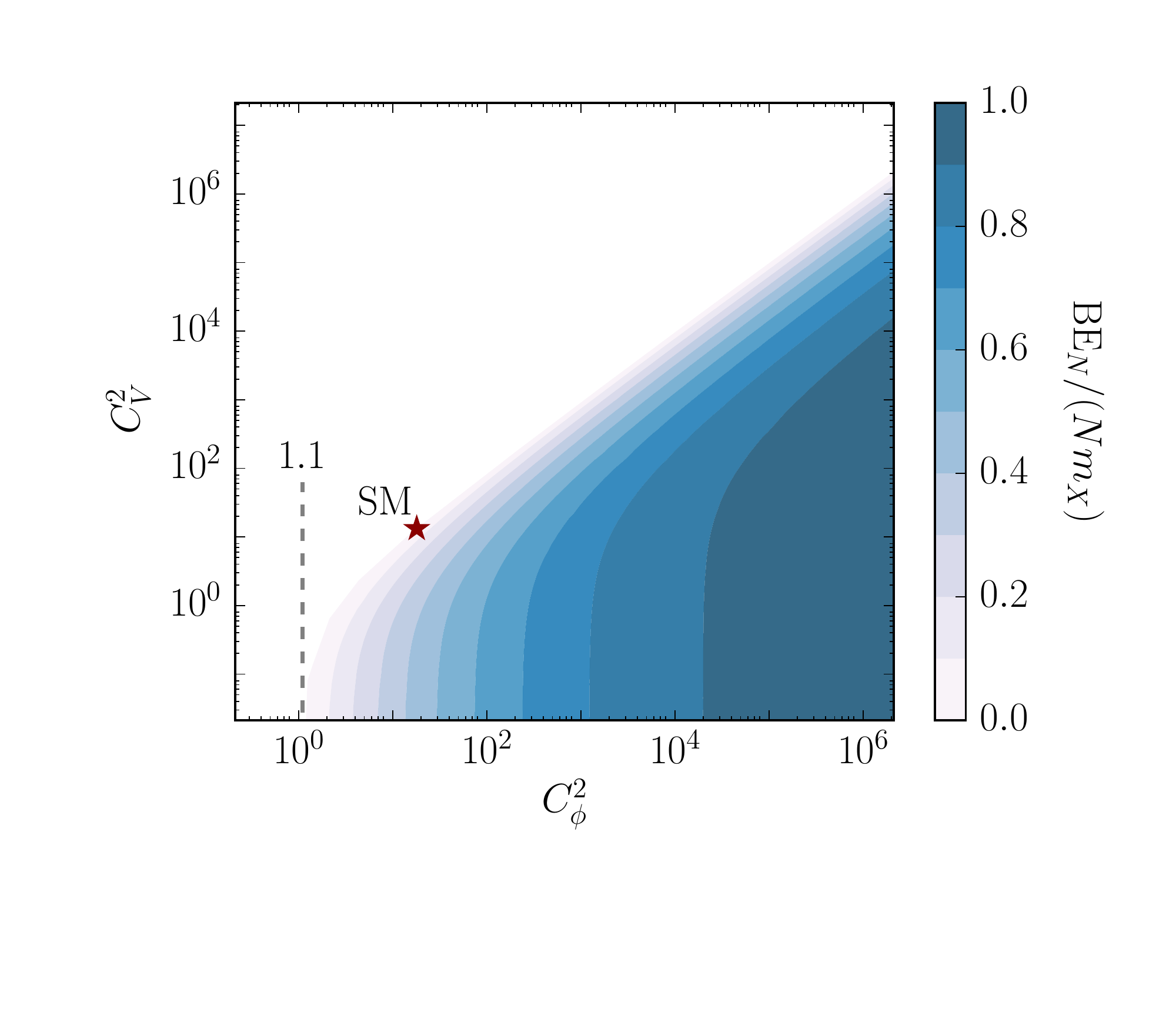}
\end{minipage}
\hspace{.3cm}
\begin{minipage}{.45\textwidth}
\includegraphics[trim={1cm 3.5cm 0cm 1cm},clip,width=1.0\linewidth]{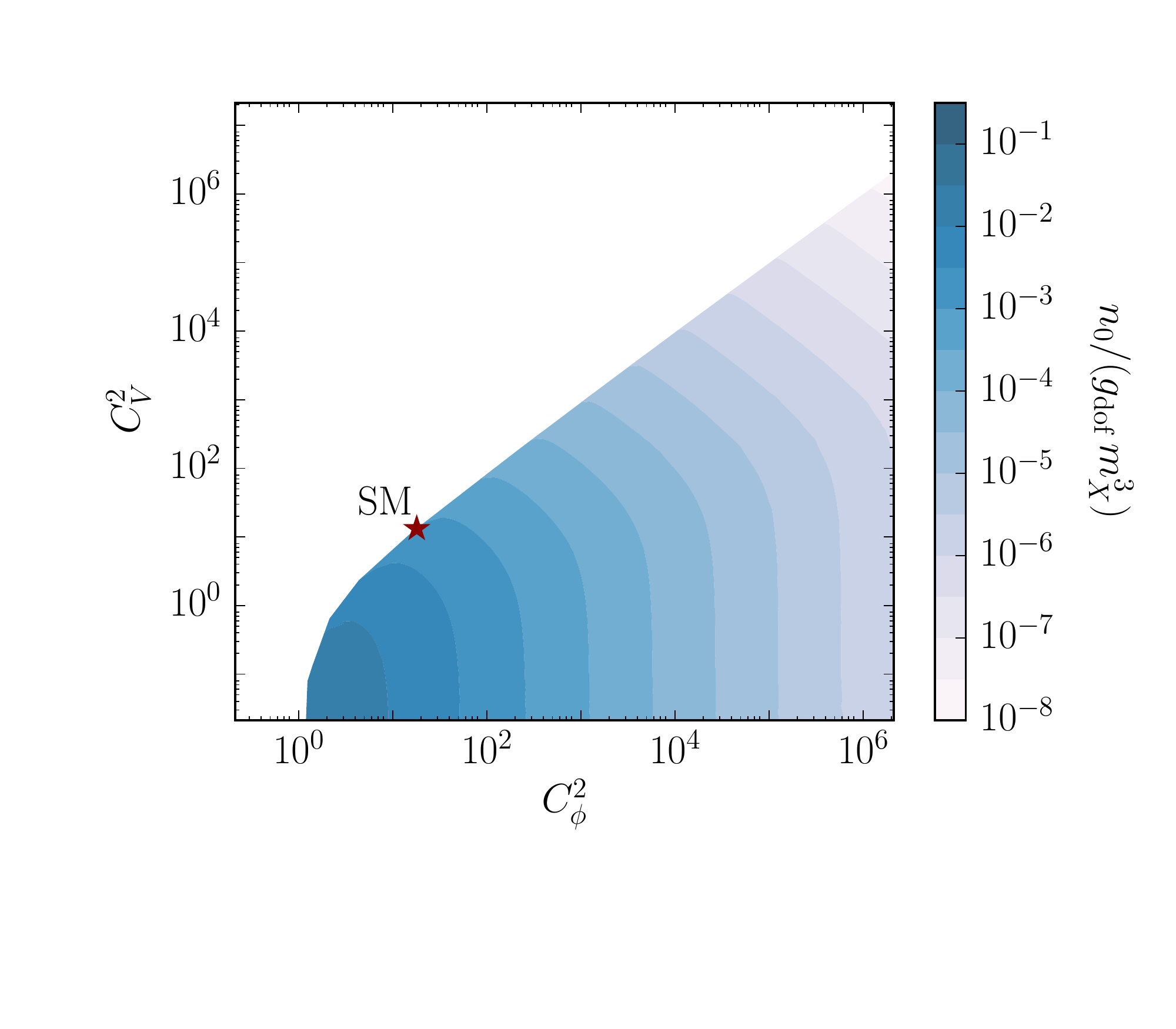}
\end{minipage}
\caption{Binding Energy (left) and the nugget density (right) as a function of the vector coupling versus scalar coupling at saturation. The SM parameters are marked as $C_\phi^2\simeq 18$ and $C_V^2 \simeq 13$. A lower bound $C_\phi^2 \gtrsim 1.1$, marked by the dashed line on the left figure, is required for saturation.}
\label{fig:binding_density}
\end{figure}

With an additional mediator, the saturation condition \Eq{eq:nugget_infinite_condition} is also modified, 
\begin{align}
1 - {m_* \over m_X} &= 3 C_\phi^2  \left( {m_* \over m_X} \right)\int_0^{k_{F0}/m_X} dx\frac{x^2}{\sqrt{x^2 + (m_*/m_X)^2}} \notag\\
p \left( {\g m_X^4 \over 6 \pi^2}\right)^{-1} &= - \frac{1}{
2 C_\phi^{2}}
\left(1 - {m_* \over m_X}\right)^2 + {1 \over 2}C_V^2 \left(\frac{k_{F0}}{m_X}\right)^6
 + \int_0^{k_{F0}/m_X} d x\, \frac{x^4}{ \sqrt{x^2 + (m_*/m_X)^2}}=0.
\label{eq:nugget_saturation}
\end{align}
where an additional parameter 
\beq C_V^2 = {2 \g \over 3 \pi} \alpha_V {m_X^2  \over  m_V^2}\,,
\eeq
is introduced. Fig~\ref{fig:binding_density} shows the binding energy in the saturation limit as a function of the vector and scalar couplings and masses, with the SM values marked by a star. The white area corresponds to $\alpha_V/m_V^2 \gtrsim \alpha_\phi/m_\phi^2$, where no infinite matter limit exists.  The lack of infinite matter limit does not necessarily imply that nuggets with saturation-like behavior do not exist. For example, in the $m_V \rightarrow 0$ limit with $\alpha_V \ll \alpha_\phi$  the vector's contribution to energy density and pressure will be a small perturbation for small enough $N$. But the impact of a massless vector grows coherently as $N^2$, and will eventually destabilize the nugget, just as Coulomb repulsion helps to destabilize large nuclei. The calculation when a light vector field is present is beyond the scope of our work, and will not be considered further here. We are mostly interested in a composite scenario mirroring nuclear matter but absent electromagnetism, where both the scalar and vector mediators are heavy, the couplings are very large, and the absence of an infinite matter (saturation) limit will generically imply the absence of large nuggets. The marked star in Fig.~\ref{fig:binding_density} shows the SM parameters, where there is a cancellation between the scalar and vector mediator such that the binding energy is small. 

\begin{figure}[h]
\centering
\includegraphics[trim={1cm 3.5cm 1cm 1cm},clip,width=.5\linewidth]{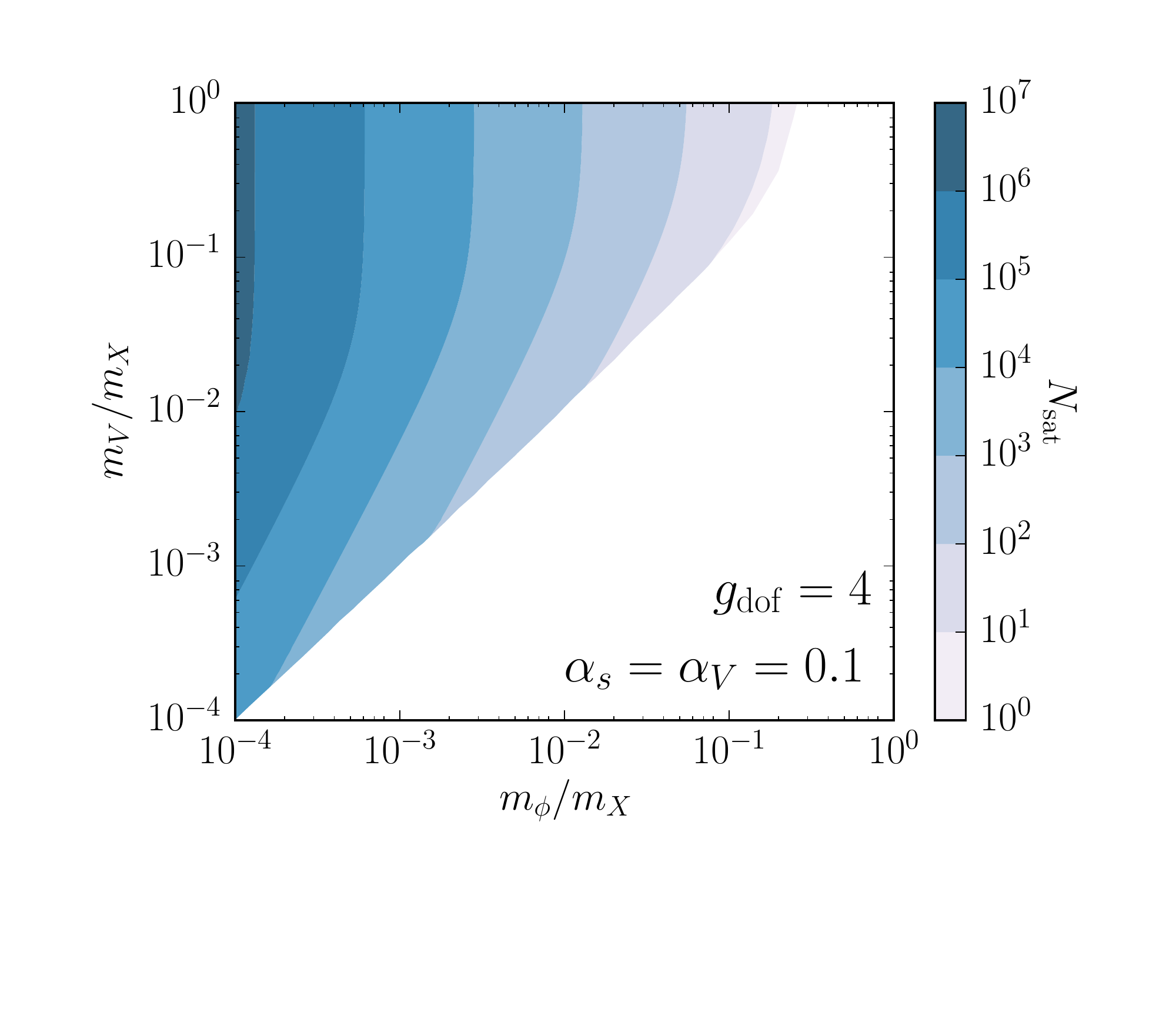}
\caption{Estimate for the lower bound $N\gtrsim N_{\rm sat}$ in order to reach saturation. The couplings are fixed $\alpha_s =\alpha_V =0.1$, and $\g=4$}
\label{fig:N_sat}
\end{figure}

For small $N$ nuggets, one also expects significant deviations from saturation.  Again defining the approach to saturation by $N_{\rm sat}=\frac{4\pi}{3}\frac{n_\text{sat}}{m_\phi^3}$,  we show in Fig.~\ref{fig:N_sat}  $N_{\rm sat}$ as a function of $m_\phi$ and $m_V$ for a benchmark with $\g=4$ and $\alpha_\phi = \alpha_V =0.1$. The inclusion of the vector mediator lowers the saturation density and accelerates the approach to saturation. 

Approximating $g_V V_0 =\left(\frac{\alpha_V\g m_X^2}{m_V^2}\right)\frac{2\g k_F^3}{3\pi m_X^2}$ (c.f.~\Eq{eq:EOM_vector}), the nugget profile is again governed by a single fermi momentum $k_F(r)$. Fig.~\ref{fig:N_sat} shows sample nugget profiles for $\alpha_\phi=\alpha_V =0.1$, with $N=100$ and three different $m_V$. One sees that increasing $m_V$ generally makes the nugget bigger and less dense, as expected.
\begin{figure}[h]
\centering
\includegraphics[trim={1cm 3.5cm 1cm 1cm},clip,width=.5\linewidth]{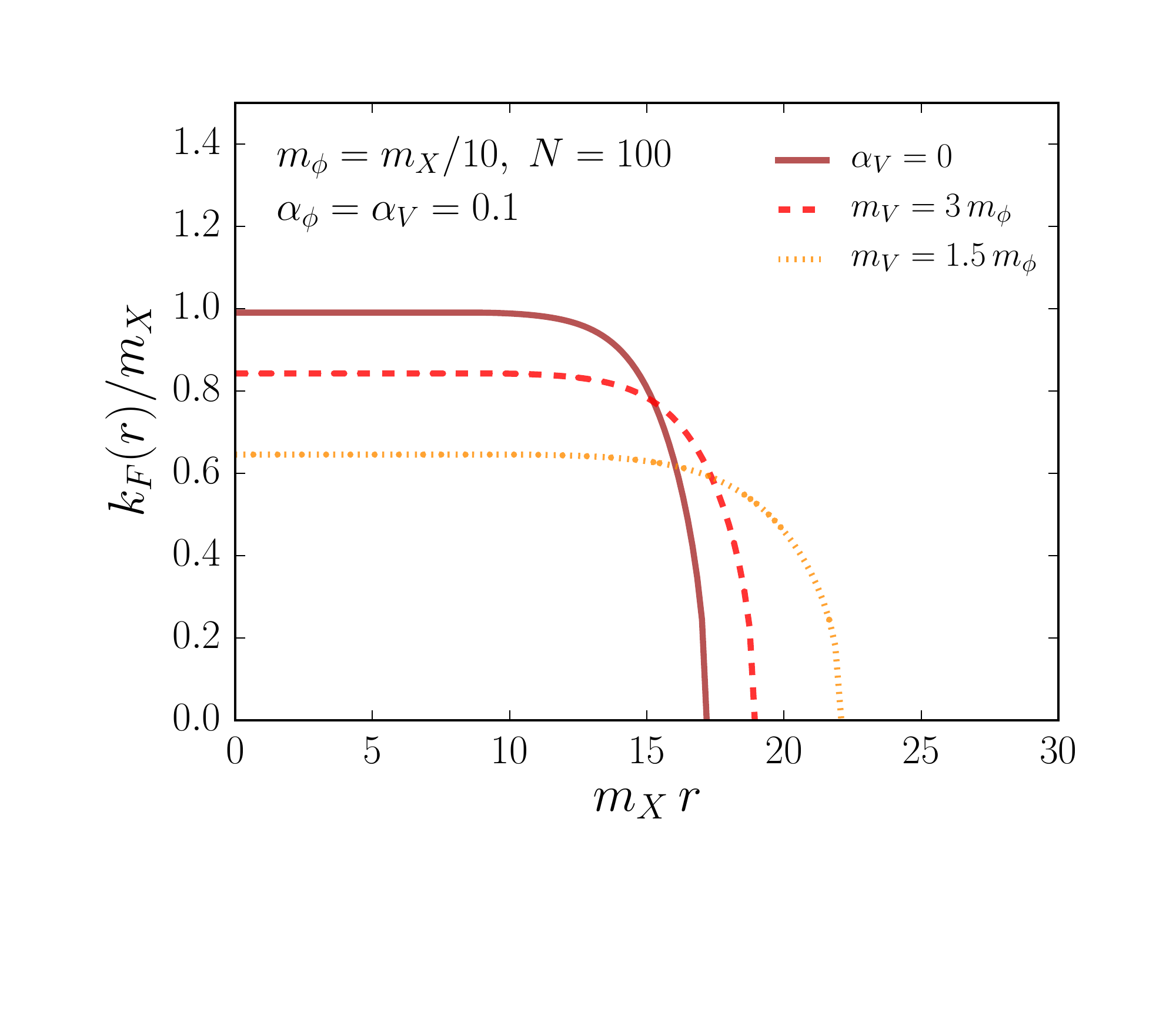}
\caption{$k_F(r)$ profile for $N=100$ nuggets with different $m_V$. The coupling is fixed at $\alpha_\phi=\alpha_V =0.1$, and $m_\phi = m_X/10$. The brown/red/orange (solid/dashed/dotted) line corresponds to the case where the vector is absent, $m_V=3 m_\phi$ and $m_V=1.5 m_\phi$ respectively}
\label{fig:kFvector}
\end{figure}

In the composite model, the nugget constituent is the dark baryon, and the vector and scalar mediators are mesons. Generically, we expect the vector and scalar mass to be comparable and of the same order as the confinement scale (with possibly a $4\pi$ suppression); the couplings should also be comparable and large. Analogous to the SM, we expect the natural parameter space to be not far from the diagonal in Fig.~\ref{fig:binding_density}. In this scenario, it is quite possible for the saturation binding energy to be a small fraction of the dark baryon mass. In this case, a new energy scale---the binding energy scale---can arise and will dictate the size and interactions of the nuggets.

\section{Conclusions}
\label{sec:conclusion}

We have studied the properties of many-particle bound states of ADM, utilizing relativistic mean field theory tools developed in nuclear physics.  The model that we applied to this system can be used, in principle, for both elementary and composite models of dark matter.  We solved the equations of motion in relativistic and non-relativistic limits, with both attractive and repulsive forces, and found a saturation property when the bound state size exceeds the force length, such that the density of the bound state nugget approaches a constant.  We found that the binding energy of these nuggets is an ${\cal O}(1)$ fraction of the rest mass, and only increases with the size of the nugget, and we derived analytic expressions for both this binding energy and the size of the nugget in the saturation limit.

Our ultimate goal is to understand the cosmology of these many-particle bound states of ADM---to determine their abundance in the Universe today and their impact on the evolution of structure in the dark sector.  With the properties of these ADM bound states in hand, written in terms of the degrees of freedom of a fundamental Lagrangian, we can follow their evolution through early universe synthesis, and ultimately through late universe structure formation.  This is the subject of our next papers. 

\section*{Acknowledgments}
We thank Dorota Grabowska, Tom Melia, Mark Wise, and Yue Zhang for comments on the draft and Martin Savage for discussions.
 HL and KZ are supported by the DoE under contract DE-AC02-05CH11231. HL is also supported by NSF grant PHY-1316783. MG is supported by the Murdock Charitable Trust.

\appendix
\section{Nugget Profile Computation}
\label{app:numerics}
Here we discuss difficulties in solving \Eq{eq:nugget_ODE} and numerical methods for circumventing them. To begin, it is convenient to rewrite \Eq{eq:nugget_ODE} in terms of $f\equiv (m_X - g_\phi \phi(r))/\mu$, which leads to
\begin{align}
\frac{1}{r}  \frac{d^2}{dr^2}(r f)= - m_\phi^2 \left(\frac{m_X}{\mu} -f\right) + 
\frac{ \mu^2\alpha_\phi\g}{2} \left[f\sqrt{1-f^2}+f^3 \log\left( \frac{f}{1+\sqrt{1-f^2}}\right) \right]
\,,
\label{eq:app_ode}
\end{align}
subject to the boundary conditions, 
\begin{align}
  f'(0)=0 \qquad
  f(R)=1 \qquad
  f'(R)=\frac{(1+m_\phi R)(m_X -\mu)}{\mu R}\,.
\label{eq:app_bdy}
\end{align}
\Eq{eq:app_ode} is highly nonlinear, but can be solved using standard numerical techniques. The nontrivial problem is to determine the values $(R,\mu)$, which are not known a priori. Solving the ODE  self-consistently with all three boundary conditions amounts to finding a curve in the two-dimensional plane, $(R(N), \mu(N))$, which can be parameterized by $N$, the nugget number, which is calculated after the fact through $N = \int d^3 \vec{r} {\g \over 6 \pi^2} k_F^3 = {\g \over 6 \pi^2} \mu^3 \int d^3 \vec{r}  \left( 1 - f^2\right)^{3/2} $. Naively, one can fix either $\mu$ or $R$, and then scan over the other variable to solve the ODE backward at $r=R$ to $r=0$. A solution is obtained when $f'(0)=0$. However, as there is a singularity at $r=0$, numerical instability can arise here. One can instead solve the ODE backward to some $r_0$ such that $f'(r_0)=0$, and find the smallest $r_0$ such that this is possible. This approach generally works well for small $N$ away from saturation. 

Near saturation, another problem emerges. Close to the saturation value ${f_0=(m_X - g_\phi \phi_0)/\mu_0}$, the right-hand-side of \Eq{eq:app_ode} approaches zero, and $|f-f_0|$ will often be exponentially small, leading to very large numerical inaccuracies. This behavior is generally expected at large $N$, which is important to solve properly in order to get accurate corrections to the saturation limit. In this regime, it is more fruitful to consider a reparametrization of the solution space, mapping $(R,\mu) \rightarrow (f(0),\mu)$. Then, fixing $\mu$, one can scan over different values of $f(0)$ and solve the ODE until $r=R$ such that $f(R)=1$, and check whether $f'(R)$ satisfies the last boundary condition.\footnote{This is essentially the strategy advocated for in \cite{Negele:1986bp}. Keep in mind that nuclei quickly reach saturation.} One immediate issue is the singularity at $r=0$, which may be resolved in the near saturation regime. Approximating \Eq{eq:app_ode}, one has
\begin{align}
  \frac{d^2}{dr^2}(r f) \simeq r \left[\kappa^2 (f-f_{0'})\right]\,,
\label{eq:ode_approx}
\end{align}
where we have performed a Taylor expansion around the zero on the right-hand-side of \Eq{eq:app_ode}. $f_{0'}$ and $\kappa^2$  both depend on $\mu$, and $\kappa^2$ is always positive as long as a saturation limit exists. When $\mu$ is close to $\mu_0$, $f_{0'}$ becomes very close to $f_0$ as well. The linearized ODE in \Eq{eq:ode_approx} can be readily solved to obtain (assuming $f'(0)=0$)
\begin{align}
  f(r)\simeq f_{0'} + \frac{\left[f(0)-f_{0'}\right]\sinh(\kappa r)}{\kappa r}\,.
\label{eq:app_sol}
\end{align}
Then, using the solution in \Eq{eq:app_sol}, one can replace the boundary condition at $r=0$ by the ones at some intermediate value $r=r_0$, where $r_0 $ is of order $1/\kappa$. It is worth noting that at very large nugget number, $|f(r)-f_{0'}|$ will typically be extremely small, and it may be useful to change variable to $l(r) \equiv -\log(f - f_0)$. It is worth noting that for the massless mediator limit, $f_0 \rightarrow 0$. For this special case, and in the large nugget limit, $f_{0'} \rightarrow 0$, and $\kappa\rightarrow\mu \sqrt{\alpha_\phi\g/2}$. The nugget profile near the origin becomes simply
\begin{align}
  f(r)\simeq \frac{f(0)\sinh(\kappa r)}{\kappa r}\,.
\label{eq:app_sol2}
\end{align}
Note that even though no saturation limit exists, for very small $\mu$, $f(r)$ will stay roughly constant as long as $r\ll 1/\mu$.

\bibliography{ADMNuggetsMerged,TL}

\end{document}